\documentclass[journal]{IEEEtran}

\ifCLASSINFOpdf

\else

\fi

\usepackage{graphicx}
\usepackage{amsthm,amssymb,amsfonts}
\usepackage{epsfig} 
\usepackage{epstopdf}
\usepackage{amsmath}
\usepackage{amsthm}
\usepackage{comment}
\usepackage{url}
\usepackage{algorithm}
\usepackage{algorithmic}
\usepackage{subfigure}
\usepackage{booktabs}

\usepackage{amsfonts}
\usepackage{bm}
\usepackage{color}
\usepackage{xcolor}
\usepackage{cite}
\usepackage{stfloats}
\usepackage{chemarrow}
\usepackage{extarrows}
\usepackage{setspace}

\usepackage{microtype}
\usepackage[american]{babel}

\usepackage[explicit]{titlesec}
\titlespacing{\section}{0pt}{1.2ex plus .0ex minus .0ex}{.3ex plus .0ex}
\titlespacing{\subsection}{0pt}{1.2ex plus .0ex minus .0ex}{.3ex plus .0ex}

\allowdisplaybreaks

\newboolean{showcomments}
\setboolean{showcomments}{true}
\newcommand{\wang}[1]{\ifthenelse{\boolean{showcomments}}
	{ \textcolor[rgb]{1,0,1}{(ZW:  #1)}}{}}
\newcommand{\fliu}[1]{\ifthenelse{\boolean{showcomments}}
	{ \textcolor{blue}{(FL:  #1)}}{}}
\newcommand{\ychen}[1]{\ifthenelse{\boolean{showcomments}}
	{ \textcolor{green}{(ZP:  #1)}}{}}
\newcommand{\slow}[1]{\ifthenelse{\boolean{showcomments}}
	{ \textcolor{blue}{(SL:  #1)}}{}}

\theoremstyle{definition}

\theoremstyle{definition}

\def\BibTeX{{\rm B\kern-.05em{\sc i\kern-.025em b}\kern-.08em
		T\kern-.1667em\lower.7ex\hbox{E}\kern-.125emX}}

\title{Distributed Emergency Frequency Control Considering Transient Stability Constraints in Multi-Infeed Hybrid AC-DC System}

\begin{document}
\setstretch{1}
\author{
	Ye~Liu,~\IEEEmembership{Student Member, IEEE}, 
	Chen~Shen,~\IEEEmembership{Senior Member, IEEE},
	and Zhaojian~Wang,~\IEEEmembership{Member, IEEE}, 
        \thanks{This work was supported by the National Key R\&D Program of China (2021YFB2400800). (Corresponding author: Chen Shen.)}     

		\thanks{Ye Liu and Chen Shen are with the State Key Laboratory of Power Systems, Department of Electrical Engineering, Tsinghua University, Beijing 100084, China (e-mail: liuye18@mails.tsinghua.edu.cn; shenchen@mail.tsinghua.edu.cn).}
		
		\thanks{Zhaojian Wang is with the Key Laboratory of System Control, and Information Processing, Ministry of Education of China, Department of Automation, Shanghai Jiao Tong University, Shanghai 200240, China (e-mail: wangzhaojian@sjtu.edu.cn).}
			
}

\markboth{Journal of \LaTeX\ Class Files,~Vol.~xx, No.~xx, xx~xxxx}%
{Shell \MakeLowercase{\textit{et al.}}: Bare Demo of IEEEtran.cls for IEEE Journals}

\maketitle

\begin{abstract}
    Due to possible emergency faults and frequency regulation reserve shortage in the multi-infeed hybrid AC-DC (MIDC) system, the emergency frequency control (EFC) with LCC-HVDC systems participating is important for system frequency stability. Nevertheless, the existing decentralized EFC strategies cannot guarantee the transient stability constraints of lines and cannot meet the frequency restoration requirement. To fill this gap, this paper proposes a complementary distributed EFC strategy. By selecting the semi-distributed or fully-distributed control law according to the state of AC system control center, the proposed distributed EFC can improve the control performance with a normal control center and enhance the control reliability in case of control center failure. Then, to derive the semi-distributed and fully-distributed control laws, a general design method is proposed by formulating an optimal EFC problem. Both of these two control laws can guarantee transient stability constraints, restore system frequency and achieve the defined optimal control objective. Moreover, the optimality of the closed-loop equilibrium is proven, and a Lyapunov analysis shows the asymptotic stability of the closed-loop system with the discontinuous control dynamics. A case study verifies the effectiveness of the proposed distributed EFC strategy.  
\end{abstract}

\begin{IEEEkeywords}
	Multi-infeed hybrid AC-DC system, distributed emergency frequency control, transient stability constraints, LCC-HVDC system.
\end{IEEEkeywords}


\section{Introduction}

\subsection{Motivation}

Multi-infeed hybrid AC-DC (MIDC) systems widely exist in the power transmission networks nowadays \cite{zhou2016china,wang2013harmonizing,karawita2009multi}, in which one AC system is connected with multiple line-commutated-converter-based HVDC (LCC-HVDC) systems. For instance, in China, more than ten LCC-HVDC systems feed in the East China regional grid. The frequency stability of the MIDC system faces many challenges. On the one hand, emergency faults (e.g., HVDC block faults) with considerable power imbalances are prone to occur; on the other hand, the feeding of LCC-HVDC systems and the asynchronous connections among regional AC systems might cause frequency regulation reserve shortage and system inertia deficiency \cite{liu2017design,bevrani2009robust}. Therefore, traditional frequency regulation strategies may not ensure the frequency stability of the MIDC system, and the emergency frequency control (EFC) strategies are urgently required. 

Due to the severe economic losses caused by the conventional EFC strategies (i.e., generator tripping and load shedding operations \cite{song2016review,terzija2006adaptive}), the more economic and effective EFC strategies considering LCC-HVDC systems' participation have been proposed in existing studies. For example, in our previous work \cite{liu2021optimal}, a decentralized EFC strategy is proposed, where the short-term overload capability of LCC-HVDC \cite{aidong2007study} is utilized to provide emergency power support. The objective of existing EFC strategies is to stabilize the system frequency when an emergency fault occurs in the MIDC system, like that of the primary frequency regulation. Hence, the existing EFC strategies will cause a steady-state frequency deviation from the nominal value. Nevertheless, considering the application in engineering practice, the following two requirements need to be met in the design of the EFC strategy:  

\textit{1) Transient Stability Constraints}. In the EFC, both the power imbalance and the power regulation amounts of the LCC-HVDC systems are considerable. Thus, the post-fault steady-state transmission power of the AC transmission lines is prone to exceed the transient stability limits \cite{sadeghzadeh1998improvement}, which might lead to transient instability problems. Therefore, the transient stability constraints of transmission lines should be considered in the design of the EFC strategy.

\textit{2) Frequency Restoration Requirement}. The system recovery after an emergency fault takes a relatively long time. For instance, when an HVDC block fault occurs, the LCC-HVDC system will not resume normal operation until about thirty minutes after fault clearing. It is inappropriate that the system frequency remains a non-nominal value during this period of time. Therefore, the EFC strategy in the MIDC system should be able to restore the system frequency. 

Since the existing EFC strategies cannot meet the above two requirements, to fill this gap, considering that the LCC-HVDC systems and the AC system are operated by different control agents, this paper proposes a novel distributed EFC strategy with the cooperative participation of LCC-HVDC systems and generators. The proposed distributed EFC strategy can guarantee the transient stability constraints of transmission lines, restore the system frequency and achieve the designed optimal control objective. 

\subsection{Literature Review}

To the best of our knowledge, there is no relevant research on considering the transient stability constraints in the EFC design. However, when the transient stability limits of transmission lines are given by the dispatch center in advance, the transient stability constraints can be regarded as the transmission power constraints. Thus, in this literature review, we focus on the frequency control strategies considering transmission power constraints of lines, which can be divided into three categories. The first category is the sensitivity-based strategy. In \cite{liu2016cooperative}, a load-side frequency control strategy considering transmission power constraints is proposed, where the controllers that might cause transmission power limits exceeded are blocked according to the sensitivity analysis. In \cite{deng2017automatic}, an automatic generation control (AGC) strategy considering security constraints of tie-lines is proposed for the wind power integrated power system, where the sensitivity coefficients are utilized to describe the impact of the wind power fluctuations on the transmission power of tie-lines. Moreover, a sensitivity-based preventive and remedial control is proposed in \cite{zhang2005implementation} to ensure that the transmission power of tie-lines does not exceed the limits during the AGC. In the above strategies, the sensitivity coefficients are related to the system parameters; however, the topology and parameters of the power system might change in case of emergency faults, and it is difficult for the system operator to accurately calculate the sensitivity coefficients in real time. Therefore, the sensitivity-based strategies \cite{liu2016cooperative,deng2017automatic,zhang2005implementation} are inapplicable to the EFC design.

The second category is the centralized strategy with transmission power constraints. In \cite{xing2011acontrol,chen2011acontrol}, a centralized active power control method is proposed for the wind-thermal integrated system to guarantee the transmission power constraints of the sending-out line. In \cite{chintam2018real}, a satin bowerbird optimization algorithm-based tertiary frequency control strategy is proposed for congestion management, which is essentially an optimal power flow (OPF)-based strategy. A real-time hybrid optimization algorithm-based congestion management strategy is proposed in \cite{esfahani2017adaptive}, which requires advanced communication system. The above centralized approaches need a control center that gathers the global information, perform centralized calculations and give control commands; due to the control center dependence and possible single-point failures, the centralized strategies might not work normally in emergency situations. Moreover, in the MIDC system, since the LCC-HVDC systems and the AC system are operated by different control agents and there is no global control center, the centralized strategies are inapplicable. 

The third category is the distributed strategy considering transmission power constraints. In \cite{wang2019distributed1,wang2019distributed2}, a distributed frequency control considering operational constraints (i.e., hard capacity constraints and tie-line power constraints) is proposed. In \cite{stegink2015port,stegink2016unifying}, a port-Hamiltonian approach is utilized to design the optimal fully-distributed frequency regulation strategy, with tie-line power constraints considered. In the above fully-distributed strategies, only the tie-line power constraints are considered; nevertheless, in the EFC of the MIDC system, besides tie-lines, other types of lines (e.g., lines near LCC-HVDC connected buses) also face the risk that the transmission power exceeds the transient stability limits, which are supposed to considered in the EFC design. In addition, in the MIDC system, the AC system usually has a control center which can control its frequency regulation units (e.g., generators) in a centralized way \cite{kundur1994power}. Thus, due to the extra communication burden and iteration dynamics of the fully-distributed control, the AC system might not be willing to adopt the fully-distributed frequency control when the control center is normal. 
    
\subsection{Contribution}

According to the literature review, to design an EFC strategy with transient stability constraints considered, the major challenges are how to make the EFC strategy suitable for the control structure of the MIDC system, and how to enhance the control reliability in emergencies. In this paper, a complementary distributed control design idea is utilized to address these challenges, and the specific control laws are derived by formulating an optimal EFC problem. The contributions of this paper are summarized as follows:
\begin{itemize}
	\item A complementary distributed EFC strategy considering transient stability constraints is proposed for MIDC systems. By selecting the semi-distributed or fully-distributed control law according to the state of AC system control center, this control strategy is suitable for the control structure of MIDC systems, can improve the control performance with a normal control center and enhance the control reliability in case of control center failure. 
	\item A general design method for both the semi-distributed and fully-distributed control laws is proposed by formulating the optimal EFC problem. Both of the two control laws can guarantee transient stability constraints, restore system frequency and achieve the defined optimal control objective. Furthermore, the implementation method of the proposed strategy in MIDC systems is introduced.   
	\item Two properties of the proposed distributed EFC strategy are analyzed. First, the optimality of the closed-loop equilibrium is proven rigorously. Second, considering the discontinuous control dynamics caused by the transient stability constraints, the asymptotic stability of the closed-loop system is proven by the Lyapunov approach. 
\end{itemize}

\subsection{Organization}

The remainder of this paper is organized as follows. Section II introduces the state model of the MIDC system. Section III proposes the complementary distributed EFC strategy, presents the fully-distributed and semi-distributed control laws, and introduces the implementation method of the proposed strategy. Section IV discusses the properties of the distributed EFC. Simulation results of an MIDC test system are given in Section V. Section VI provides the conclusion.    

\section{State Model of MIDC System}

\begin{figure}[tb]
	\centering
	\includegraphics[width=0.43\textwidth]{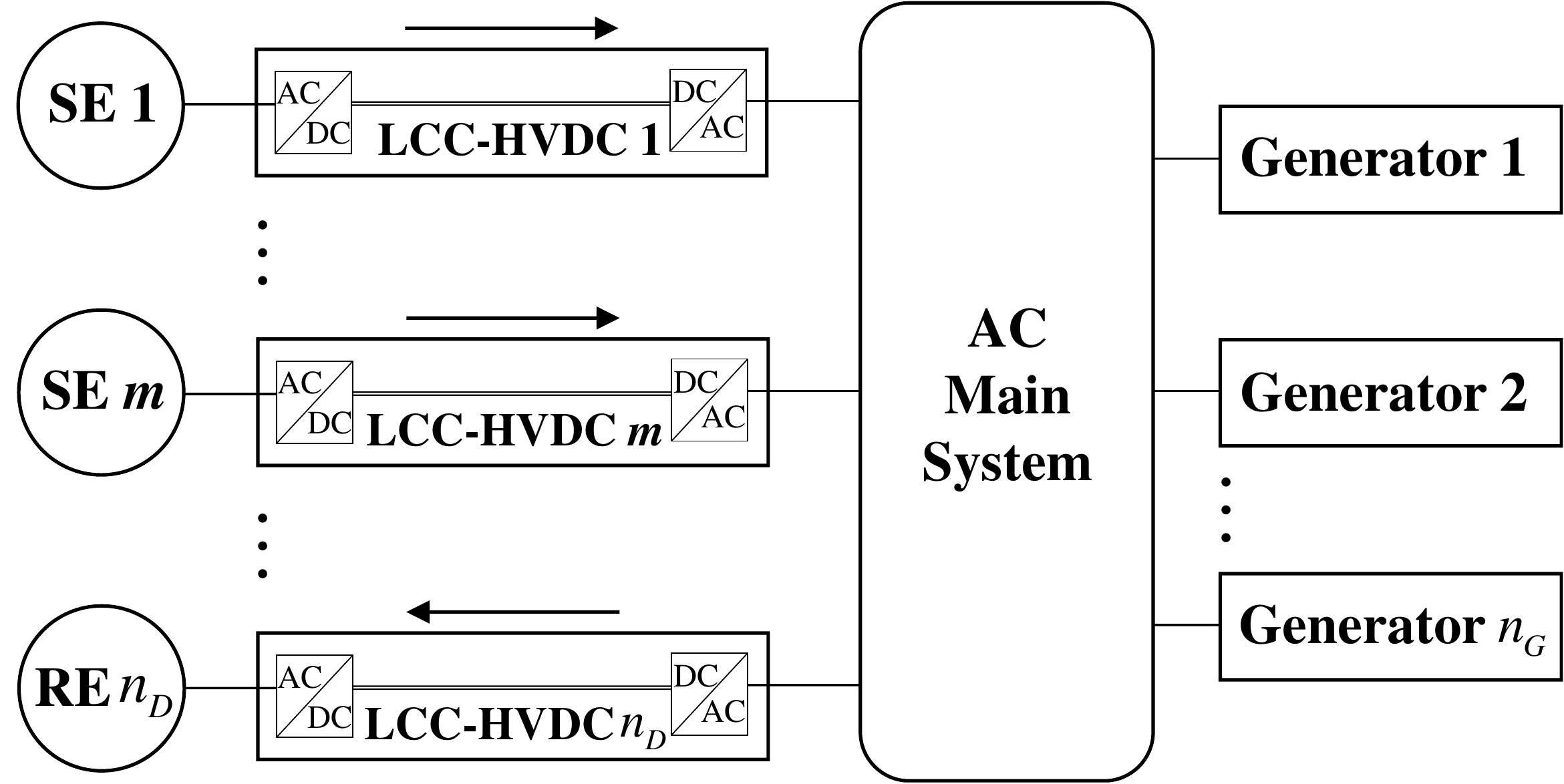}
	\caption{Topology of the MIDC system.}
	\label{topo_midc}
\end{figure}

The topology of an MIDC system is shown in Fig. \ref{topo_midc}. In this MIDC system, the AC main system contains $n_G$ generators and is connected with $n_D$ LCC-HVDC systems. There are $m$ LCC-HVDC systems transmitting power from sending-end (SE) systems to the AC main system while ($n_D - m$) LCC-HVDC systems from the AC main system to receiving-end (RE) systems, which are called SE-LCC and RE-LCC systems respectively. The SE 1$\sim$$m$ and RE ($m$+1)$\sim$$n_D$ systems are collectively called the adjacent AC systems. Since this paper focuses on the EFC for the AC main system, we ignore the dynamics of the adjacent AC systems in the state modeling. 

From the perspective of graph theory, the AC main system can be described by a directed connected graph $\mathcal{G}=(\mathcal{N},\mathcal{E})$, where $\mathcal{N}$ represents the buses and $\mathcal{E} \subseteq \mathcal{N} \times \mathcal{N}$ represents the transmission lines. The buses can be divided into three types, i.e., the generator buses, the LCC-HVDC connected buses and the passive load buses, which are denoted by $\mathcal{N}_G$, $\mathcal{N}_D$ and $\mathcal{N}_P$. We have $\mathcal{N}=\mathcal{N}_G \cup \mathcal{N}_D \cup \mathcal{N}_P$. The numbers of buses in the above sets are denoted by $n$, $n_G$, $n_D$ and $n_P$, and $n=n_G +n_D +n_P$. A transmission line in $\mathcal{E}$ can be represented by either $e\in \mathcal{E}$ or $ij \in\mathcal{E}$. Since graph $\mathcal{G}$ is connected, if $ij \in \mathcal{E}$, then $ji \notin \mathcal{E}$. The number of transmission lines is denoted by $n_e$. Define the set of neighboring buses of bus $i$ as $\mathcal{N}_i = \{j\in \mathcal{N} \big| ij\in \mathcal{E}$ or $ji\in \mathcal{E}\}$. We have the following assumptions.

\textbf{Assumption 1:} 1) A bus in $\mathcal{N}_D$ is only connected with one LCC-HVDC system. 2) An LCC-HVDC connected bus cannot be a generator bus, i.e., $\mathcal{N}_D \cap \mathcal{N}_G =\varnothing$. 3) The DC power flow is adopted in the modeling, which is reasonable for transmission networks.  

Base on the above assumptions, the state model of the MIDC system in Fig. \ref{topo_midc} is described by the differential algebraic
equations (DAEs) in \eqref{system_model_dae}, where the second-order rotation model of generators and the first-order power regulation inertia model of generators and LCC-HVDC systems are considered \cite{kundur1994power}.
\begin{subequations}
	\label{system_model_dae}
	\begin{align}
	&\dot{\theta}_i=\omega_i,\ i \in \mathcal{N}
	\label{system_model_dae_1}\\
	&M_i \dot{\omega}_i +D_i \omega_i = P_i^{in} + p_i^G - \sum_{e\in \mathcal{E}} C_{i,e} P_e,\ i \in \mathcal{N}_G
	\label{system_model_dae_2}\\
	&0= P_i^{in} + p_i^D - \sum_{e\in \mathcal{E}} C_{i,e} P_e,\ i \in \mathcal{N}_D
	\label{system_model_dae_3}\\
	&0= P_i^{in} - \sum_{e\in \mathcal{E}} C_{i,e} P_e,\ i \in \mathcal{N}_P\label{system_model_dae_4}\\
	&T_i^G \dot{p}_i^G = -p_i^G+P_i^G + u_i^G ,\ i \in \mathcal{N}_G \label{system_model_dae_5}\\
	&T_i^D \dot{p}_i^D = -p_i^D+P_i^D + u_i^D,\ i \in \mathcal{N}_D \label{system_model_dae_6}
	\end{align}
\end{subequations}
where $\theta_i$ is the phase angle with reference to the synchronous rotation coordinate, $\omega_i$ is the frequency deviation from the nominal value, $M_i>0$ and $D_i>0$ are respectively the inertia time constant and the damping constant of generator $i$, $P_i^{in}$ is the total power injection ($>0$) or demand ($<0$) excluding generators and LCC-HVDC systems, $p_i^G$ is the active power output of generator $i$, $p_i^D$ is the transmission power of LCC-HVDC $i$ which is positive for SE-LCC or negative for RE-LCC, $P_i^G$ and $P_i^D$ are the scheduled values of $p_i^G$ and $p_i^D$, $u_i^G$ and $u_i^D$ are the control orders for generator $i$ and LCC-HVDC $i$ respectively, $T_i^G$ and $T_i^D$ are the power regulation inertia constants, $P_e$ is the transmission power of line $e$, and $C_{i,e}$ is the element of the $n\times n_e$ node-branch incident matrix $C$, where $C_{i,e}=1$ if $e=ij\in \mathcal{E}$, $C_{i,e}=-1$ if $e=ji\in \mathcal{E}$, and $C_{i,e}=0$ otherwise. Under the DC power flow assumption, the transmission power of line $e=ij$ is represented as:
\begin{align}
P_e = P_{ij} = B_{ij}(\theta_i - \theta_j)
\label{p_e_ij}
\end{align}
where $B_{ij}$ is the susceptance of line $ij$. Under the definition of $P_{ij}$, \eqref{system_model_dae_1} can be equivalently represented as:
\begin{align}
\label{system_model_dae_1_1}
\dot{P}_{ij} = B_{ij} (\omega_i - \omega_j),\ ij\in\mathcal{E}
\end{align}

In practical engineering, the transient stability constraints of transmission lines are usually derived in advance by the off-line calculations at the grid dispatch center. Therefore, the transient stability constraints in the control design can be represented as the transmission power constraints:
\begin{align}
\underline{P}_{ij} \le P_{ij} \le \overline{P}_{ij},\ ij \in \mathcal{E}
\label{trans_cons}
\end{align}
where $\underline{P}_{ij}$ and $\overline{P}_{ij}$ are the lower and upper bounds which are given constants. The transient stability constraints are only updated when the system operation state changes. Based on the described model of the MIDC system, we further design the distributed EFC strategy. 

\section{Distributed Emergency Frequency Control}

In this section, we first introduce the overall framework of the complementary distributed EFC strategy. Then, with the general design method, the fully-distributed and semi-distributed control laws are derived respectively by formulating the optimal EFC problem, and the implementation method of the proposed strategy in MIDC systems is presented.

\subsection{Overall Control Framework}

In the proposed distributed EFC strategy, the generators and LCC-HVDC systems cooperatively participate in the EFC to stabilize the system frequency, further restore the system frequency and guarantee the transient stability constraints of key transmission lines. Note that the LCC-HVDC systems only provide power support in case of emergency faults, and maintain scheduled constant transmission power during slight frequency fluctuations. The short-term overload capability of LCC-HVDC systems is utilized to provide emergency power support \cite{aidong2007study}. In the existing power grid structure, the AC main system usually has a control center which can control the generators in this system, while the LCC-HVDC systems are controlled by other control agents. Therefore, the distributed control design idea is utilized to adapt to the existing grid structure with multiple control agents. The distributed EFC strategy proposed in this paper contains two complementary control laws, which are applicable for different states (normal or failure) of the control center of the AC main system:

\textit{1) Semi-Distributed Control Law}. The control center of the AC main system gathers the information of buses in $\mathcal{N}_G \cup \mathcal{N}_P$ and gives control orders to the generators in a centralized way. Meanwhile, the control center, regarded as a virtual bus, conducts communications and distributed control iterations with LCC-HVDC connected buses in $\mathcal{N}_D$.

\textit{2) Fully-Distributed Control Law}. All the buses in $\mathcal{N}$ only conduct communications and distributed control iterations with neighboring buses. 

The specific semi-distributed and fully-distributed control laws are derived in Section III.D and Section III.C respectively, where the transient stability constraints of transmission lines are considered. In the subsequent analysis and case study, we will see that these two control laws have similar steady-state performance, and compared with the fully-distributed control law, the semi-distributed control law has better transient performance and has less communication burden when there are numerous transmission lines. Based on the aforementioned two control laws, the framework of the complementary distributed EFC strategy is shown in Fig. \ref{overall_framework}, which mainly includes the following three points:

\begin{figure}[htb]
	\centering
	\includegraphics[width=0.48\textwidth]{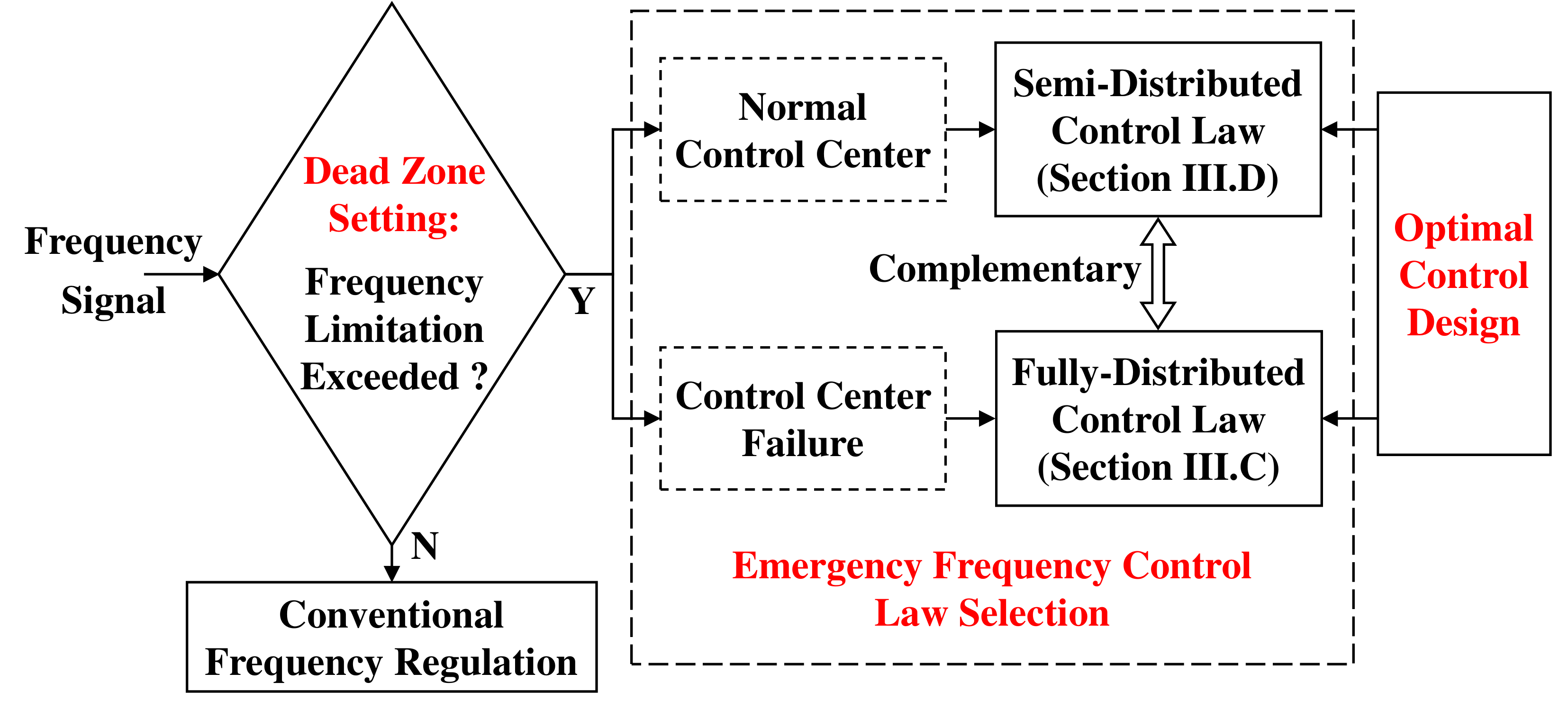}
	\caption{Overall framework of complementary distributed EFC.}
	\label{overall_framework}
\end{figure}

\textit{1) Dead Zone Setting}. Since the distributed EFC strategy should only work when emergency faults occur, a dead zone setting is required to determine whether there is an emergency fault and whether to enable the distributed EFC. In this paper, the frequency deviation limitation is utilized to set the dead zone. If the system frequency is within the limitation, only the conventional generator-based frequency regulation strategy works; if the system frequency exceeds the limitation, the distributed EFC strategy is enabled. Considering the determination of the frequency limitation value, since the grid operator would like to give priority to the distributed EFC strategy rather than load shedding to deal with the emergency frequency drops, the limitation value of the dead zone is supposed to be higher than the threshold value of load shedding.

\textit{2) Emergency Frequency Control Law Selection}. The distributed EFC strategy selects the control law according to the state of the control center of the AC main system. When the control center works normally, the semi-distributed control law with better transient performance is selected. If the control center fails (e.g., a single-point fault caused by communication failure), the control-center-free fully-distributed control law will be selected as a backup control. Combining these two control laws, the complementary distributed EFC strategy not only improves the control performance with a normal control center, but also avoids the control center dependence and enhances the control reliability in case of control center failure. 

\textit{3) Optimal Control Design}. The optimal semi-distributed and fully-distributed control laws can achieve the defined steady-state optimization objective, with transient stability constraints guaranteed and frequency restored. In the following part of this section, to appropriately allocate the power imbalance among the generators and LCC-HVDC systems, the optimal EFC problem is formulated and the optimal control laws are derived by a general design method. 

\subsection{Optimal Emergency Frequency Control Problem with Transient Stability Constraints}

In the MIDC system, the power imbalance caused by an emergency fault can be represented as $\sum_{i\in\mathcal{N}} P_i^{in} + \sum_{i\in\mathcal{N}_G} P_i^{G} + \sum_{i\in\mathcal{N}_D} P_i^{D} \neq 0$. The optimal EFC problem with transient stability constraints (TS-OEFC) is formulated to appropriately allocate this power imbalance by minimizing the designed total control cost. The dead zone setting is not considered in the TS-OEFC problem since it has no effect on the system steady state. 

We define the cost function of generator $i$ in a classic form: $C_i^G(u_i^G) = \frac{1}{2}\alpha_i (u_i^G)^2$, where $\alpha_i$ is the cost coefficient. For LCC-HVDC $i$, a general quadratic cost function can be defined, which is applicable to various control objectives. In this paper, to appropriately allocate the power imbalance, we select the control objective as that the LCC-HVDC with a larger regulation margin provides more power support; for more control objectives, we refer to \cite{liu2021optimal}. With the selected control objective, the cost function of LCC-HVDC $i$ is represented as:
\begin{align}
C_i^D(u_i^D) = \tilde{\beta}_i ( \frac{u_i^D}{Z_i^D})^2 = \frac{\tilde{\beta}_i}{(Z_i^D)^2}(u_i^D)^2
\end{align}
where $\tilde{\beta}_i$ and $Z_i^D$ are the cost coefficient and power regulation margin of LCC-HVDC $i$, respectively. Then, define $\beta_i = \frac{2 \tilde{\beta}_i}{(Z_i^D)^2}$ is the effective cost coefficient, and the optimization objective of the TS-OEFC problem is:
\begin{align}
\min\ \sum_{i\in \mathcal{N}_G}\frac{1}{2}\alpha_i (u_i^G)^2 + \sum_{i\in \mathcal{N}_D}\frac{1}{2}\beta_i (u_i^D)^2
\label{obj_func}
\end{align} 

In this TS-OEFC problem, to consider the transient stability constraints \eqref{trans_cons} and to benefit the control design in the following subsections, we introduce a new variable for each bus in $\mathcal{N}$, i.e., the virtual phase angle $\varphi_i,i\in\mathcal{N}$. Then, the virtual power flow of line $e=ij$ is defined as:
\begin{align}
\hat{P}_e = \hat{P}_{ij} = B_{ij}(\varphi_i - \varphi_j)
\end{align}
In fact, the virtual phase angle $\varphi_i$ is a cyber variable in the designed distributed EFC, and in Theorem 1, we will prove that the virtual power flow $\hat{P}_{ij}$ equals to the actual power flow $P_{ij}$ at the optimum of the TS-OEFC problem. During the system steady state, \eqref{system_model_dae_5} and \eqref{system_model_dae_6} yield $p_i^G = P_i^G + u_i^G$ and $p_i^D = P_i^D + u_i^D$. By these variable substitutions and adding a quadratic penalty item to \eqref{obj_func}, the TS-OEFC problem is shown as follows: 
\begin{subequations}
	\label{ts_oefc}
	\begin{align}
	&\min_{p^G,p^D,\omega^G,P,\varphi} \sum_{i\in \mathcal{N}_G}\frac{1}{2}\alpha_i (p_i^G-P_i^G)^2 \label{ts_oefc_1}
	\\ \nonumber&\ \ \ \ \ \ \ \ \ \ \ \ + \sum_{i\in \mathcal{N}_D}\frac{1}{2}\beta_i (p_i^D-P_i^D)^2 + \sum_{i\in \mathcal{N}_G}\frac{1}{2}D_i \omega_i^2
	\\ \label{ts_oefc_2}&\text{s.t.}\ \ P_i^{in} + p_i^G -D_i \omega_i - \sum_{e\in \mathcal{E}} C_{i,e} P_e = 0,\ i \in \mathcal{N}_G
	\\ \label{ts_oefc_3}&\ \ \ \ \ P_i^{in} + p_i^D - \sum_{e\in \mathcal{E}} C_{i,e} P_e=0,\ i \in \mathcal{N}_D
	\\ \label{ts_oefc_4}&\ \ \ \ \ P_i^{in} - \sum_{e\in \mathcal{E}} C_{i,e} P_e=0,\ i \in \mathcal{N}_P
	\\ \label{ts_oefc_5}&\ \ \ \ \ P_i^{in} + p_i^G - \sum_{j\in \mathcal{N}_i} B_{ij}(\varphi_i - \varphi_j) = 0,\ i \in \mathcal{N}_G
	\\ \label{ts_oefc_6}&\ \ \ \ \ P_i^{in} + p_i^D - \sum_{j\in \mathcal{N}_i} B_{ij}(\varphi_i - \varphi_j)=0,\ i \in \mathcal{N}_D
	\\ \label{ts_oefc_7}&\ \ \ \ \ P_i^{in} - \sum_{j\in \mathcal{N}_i} B_{ij}(\varphi_i - \varphi_j)=0,\ i \in \mathcal{N}_P
	\\ \label{ts_oefc_8}&\ \ \ \ \ B_{ij}(\varphi_i - \varphi_j)-\overline{P}_{ij} \le 0,\ ij\in \mathcal{E}
	\\ \label{ts_oefc_9}&\ \ \ \ \ \underline{P}_{ij} - B_{ij}(\varphi_i - \varphi_j) \le 0,\ ij\in \mathcal{E}
	\end{align}
\end{subequations}
where the vectors $p^G = \{ p_i^G, i\in \mathcal{N}_G \}$, $p^D = \{ p_i^D, i\in \mathcal{N}_D \}$, $\omega^G = \{ \omega_i, i\in \mathcal{N}_G \}$, $P = \{ P_e, e\in \mathcal{E} \}$, $\varphi = \{ \varphi_i, i\in \mathcal{N} \}$. We make the following assumption.

\textbf{Assumption 2:} The TS-OEFC problem \eqref{ts_oefc} is feasible.

By choosing the reference angles $\theta_1$ and $\varphi_1$, bus 1$\in \mathcal{N}_G$, we define the relative phase angles $\tilde{\theta}_i = \theta_i - \theta_1$ and the relative virtual phase angles $\tilde{\varphi}_i = \varphi_i - \varphi_1$, $i\in \mathcal{N}$. We have $\tilde{\theta}_i - \tilde{\theta}_j = \theta_i - \theta_j$, $\tilde{\varphi}_i - \tilde{\varphi}_j = \varphi_i - \varphi_j$, and $\tilde{\theta}_1 = \tilde{\varphi}_1 = 0$. We denote the vectors ${\theta} = \{ {\theta}_i, i\in \mathcal{N} \}$, $\tilde{\theta} = \{ \tilde{\theta}_i, i\in \mathcal{N} \}$ and $\tilde{\varphi} = \{ \tilde{\varphi}_i, i\in \mathcal{N} \}$.

Note that \eqref{ts_oefc} is a strictly convex optimization problem. Let $\tau = \{ \tau_i,i\in\mathcal{N}\}$, $\lambda = \{ \lambda_i,i\in\mathcal{N}\}$, $\gamma^+ = \{ \gamma_{ij}^+,ij\in\mathcal{E}\}$ and $\gamma^- = \{ \gamma_{ij}^-,ij\in\mathcal{E}\}$ denote the Lagrangian multipliers of constraints \eqref{ts_oefc_2}-\eqref{ts_oefc_4}, \eqref{ts_oefc_5}-\eqref{ts_oefc_7}, \eqref{ts_oefc_8} and \eqref{ts_oefc_9} respectively. And let $x=(p^G,p^D,\omega^G,P,\varphi)$, $y=(\tau,\lambda)$ and $z=(\gamma^+,\gamma^-)$. Then, the Lagrangian function of the TS-OEFC problem can be represented as:
\begin{align}
\label{lag_0}
&L(x,y,z) = \sum_{i\in \mathcal{N}_G}\frac{1}{2}\alpha_i (p_i^G-P_i^G)^2 + \sum_{i\in \mathcal{N}_D}\frac{1}{2}\beta_i (p_i^D-P_i^D)^2 \nonumber 
\\ &+ \sum_{i\in \mathcal{N}_G}\frac{1}{2}D_i \omega_i^2 + \sum_{i\in \mathcal{N}_G} \left((\tau_i+\lambda_i)p_i^G - \tau_i D_i \omega_i \right) \nonumber
\\ & + \sum_{i\in \mathcal{N}_D} (\tau_i+\lambda_i)p_i^D + \sum_{i\in \mathcal{N}} \tau_i \bigg( P_i^{in} - \sum_{e\in \mathcal{E}} C_{i,e} P_e \bigg) \nonumber
\\& + \sum_{i\in \mathcal{N}} \lambda_i \bigg( P_i^{in} - \sum_{j\in \mathcal{N}_i} B_{ij}(\varphi_i - \varphi_j) \bigg) \nonumber
\\& + \sum_{ij \in \mathcal{E}} \gamma_{ij}^+ \left(B_{ij}(\varphi_i - \varphi_j)-\overline{P}_{ij}\right) \nonumber
\\& + \sum_{ij \in \mathcal{E}} \gamma_{ij}^- \left(\underline{P}_{ij} - B_{ij}(\varphi_i - \varphi_j)\right)
\end{align}

We have the following theorem about the TS-OEFC problem.

\textbf{Theorem 1:} If Assumption 2 holds, at the optimum of the TS-OEFC problem $x^* = (p^{G*},p^{D*},\omega^{G*},P^*,\varphi^*)$:
\begin{enumerate}
	\item the system frequency is restored, i.e., $\omega_i^* = 0,\ i\in\mathcal{N}_G$.
	\item the transient stability constraints of transmission lines are guaranteed, i.e., $\underline{P}_{ij} \le P_{ij}^* \le \overline{P}_{ij},\ ij \in \mathcal{E}$.
\end{enumerate}
\begin{proof}
	1) First, under Assumption 2, the optimum $(p^{G*},p^{D*},\omega^{G*},P^*,\varphi^*)$ satisfies the constraints \eqref{ts_oefc_2}-\eqref{ts_oefc_9} of the TS-OEFC problem. Then, summing all the equations in \eqref{ts_oefc_2}-\eqref{ts_oefc_4} at the optimum, we have:
	\begin{align}
	\label{theorem_1}
	\sum_{i\in\mathcal{N}} P_i^{in} + \sum_{i\in\mathcal{N}_G} p_i^{G*} + \sum_{i\in\mathcal{N}_D} p_i^{D*} - \sum_{i\in\mathcal{N}_G} D_i \omega_i^* =0
	\end{align}
	Similarly, summing all the equations in \eqref{ts_oefc_5}-\eqref{ts_oefc_7}, we have:
	\begin{align}
	\label{theorem_2}
	\sum_{i\in\mathcal{N}} P_i^{in} + \sum_{i\in\mathcal{N}_G} p_i^{G*} + \sum_{i\in\mathcal{N}_D} p_i^{D*} =0
	\end{align}
	Combining \eqref{theorem_1} and \eqref{theorem_2}, we derive $\sum_{i\in\mathcal{N}_G} D_i \omega_i^* = 0$.
	
	Since all the constraints of the TS-OEFC problem are linear, the Slater's condition holds. Thus, according to the KKT conditions \cite{boyd2004convex}, we have:
	\begin{subequations}
		\begin{align}
		&\frac{\partial L}{\partial \omega_i} \big| _{(x^*,y^*,z^*)}= D_i (\omega_i^* -\tau_i^*)=0,\ i\in\mathcal{N}_G
		\\ &\frac{\partial L}{\partial P_{ij}} \big| _{(x^*,y^*,z^*)}= \tau_j^*-\tau_i^*=0,\ ij\in\mathcal{E}
		\end{align}
	\end{subequations}
	which yields $\omega_i^* = \omega_j^* = \omega_{syn}$, where $\omega_{syn}$ is the steady-state synchronous frequency deviation of the power system. Thus:
	\begin{align}
	\sum_{i\in\mathcal{N}_G} D_i \omega_i^* = \omega_{syn} \sum_{i\in\mathcal{N}_G} D_i = 0
	\end{align}
	Due to $D_i >0$, we have $\omega_{syn}=\omega_i^* = 0,\ i\in \mathcal{N}_G$, which means that the system frequency is restored.
	
	2) Since $\sum_{e\in \mathcal{E}} C_{i,e} P_e = \sum_{j\in \mathcal{N}_i} B_{ij}(\theta_i - \theta_j)$, by setting $\omega_i^*=0$ in \eqref{ts_oefc_2} and comparing \eqref{ts_oefc_2}-\eqref{ts_oefc_4} and \eqref{ts_oefc_5}-\eqref{ts_oefc_7} at the optimum, we have:
	\begin{align}
	\sum_{j\in \mathcal{N}_i} B_{ij}(\theta_i^* - \theta_j^*) = \sum_{j\in \mathcal{N}_i} B_{ij}(\varphi_i^* - \varphi_j^*),\ i\in\mathcal{N}
	\label{eq_13}
	\end{align}
	Define the matrix $B=diag(B_{ij}), ij\in\mathcal{E}$, we can rewrite \eqref{eq_13} in the vector form:
	\begin{align}
	CBC^T\theta^* = CBC^T\varphi^*
	\label{eq_14}
	\end{align} 
	Note that $CBC^T$ is the $n\times n$ Laplacian matrix of graph $\mathcal{G}$ with line weights $B_{ij}$, thus, the rank of $CBC^T$ is $(n-1)$ \cite{brualdi1991combinatorial}. 
	
	Under the definition of the relative phase angles, \eqref{eq_14} yields:
	\begin{align}
	CBC^T\tilde{\theta}^* = CBC^T\tilde{\varphi}^*
	\label{eq_15}
	\end{align}
	Note that $\tilde{\theta}_1 =0$ and $ \tilde{\varphi}_1 = 0$ are constants. Hence, the actual coefficient matrix of the linear equations \eqref{eq_15} is an $(n-1)$-order principal minor of $CBC^T$ with rank $(n-1)$, and we have $\tilde{\theta}^* = \tilde{\varphi}^*$. Since $\tilde{\theta}_i - \tilde{\theta}_j = \theta_i - \theta_j$, $\tilde{\varphi}_i - \tilde{\varphi}_j = \varphi_i - \varphi_j$, we have $B_{ij} (\theta_i^* - \theta_j^*) = B_{ij} (\varphi_i^* - \varphi_j^*)$, i.e., $P_{ij}^* = \hat{P}_{ij}^*,\ ij\in \mathcal{E}$, which means that the virtual power flow equals to the actual power flow at the optimum. Further, combining constraints \eqref{ts_oefc_8}-\eqref{ts_oefc_9}, we have $\underline{P}_{ij} \le P_{ij}^* \le \overline{P}_{ij},\ ij\in \mathcal{E}$, which shows that the transient stability constraints of transmission lines are guaranteed at the optimum.
\end{proof}

Theorem 1 elaborates that the TS-OEFC problem implicitly considers the transient stability constraints of transmission lines and the frequency restoration requirement. Since $\omega_i^*=0$ at the optimum, the third item in \eqref{ts_oefc_1} is a penalty item and has no effect on the optimum of the TS-OEFC problem. Then, based on the TS-OEFC probelm, we first design the fully-distributed control law, and then elaborate how to derive the semi-distributed control law by the general design method.
 
\subsection{Fully-Distributed Control Law}

The optimal fully-distributed control law of the EFC strategy is designed as follows:
\begin{subequations}
	\label{fully-distri}
	\begin{align}
	&u_i^G = -\frac{1}{\alpha_i}(\omega_i+\lambda_i),\ i\in\mathcal{N}_G \label{fully-distri_1}
	\\ &u_i^D = -\frac{1}{\beta_i}(\omega_i+\lambda_i),\ i\in\mathcal{N}_D \label{fully-distri_2}
	\\ &\dot{\lambda}_i = K_i^{\lambda}\bigg( d_i (M_i \dot{\omega_i}+D_i \omega_i) + \sum_{j\in \mathcal{N}_i} B_{ij}(\theta_i - \theta_j)\nonumber 
	\\&\ \ \ \ \ \ \ \ \ \ \ \ \ \  - \sum_{j\in \mathcal{N}_i} B_{ij}(\varphi_i - \varphi_j) \bigg),\ i\in\mathcal{N} \label{fully-distri_3}
	\\&\dot{\varphi}_i = K_i^{\varphi} \bigg( \sum_{j\in \mathcal{N}_i} B_{ij}(\lambda_i - \lambda_j) \nonumber
	\\&\ \ \ \ \ \ \ \ \ \ \ \ \ \  - \sum_{e\in \mathcal{E}} C_{i,e} B_e (\gamma_e^+ - \gamma_e^-) \bigg),\ i\in\mathcal{N} \label{fully-distri_4}
	\\&\dot{\gamma}_{ij}^+ = K_{ij}^{\gamma^+} \left[B_{ij} (\varphi_i - \varphi_j)-\overline{P}_{ij} \right]_{\gamma_{ij}^+}^+,\ ij\in\mathcal{E} \label{fully-distri_5}
	\\&\dot{\gamma}_{ij}^- = K_{ij}^{\gamma^-} \left[\underline{P}_{ij} - B_{ij} (\varphi_i - \varphi_j) \right]_{\gamma_{ij}^-}^+,\ ij\in\mathcal{E} \label{fully-distri_6}
	\end{align}
\end{subequations}
where $K_i^{\lambda}$, $K_i^{\varphi}$, $K_{ij}^{\gamma^+}$ and $K_{ij}^{\gamma^-}$ are the positive control parameters, $d_i = 1$ for $i\in \mathcal{N}_G$ while $d_i = 0$ for $i\in \mathcal{N}_D \cup \mathcal{N}_P$, and the projection operator $[b]_a^+$ in \eqref{fully-distri_5}-\eqref{fully-distri_6} is defined as:
\begin{align}
\label{define_ba}
[b]_a^+ = \left\{\begin{array}{l} b,\ a>0\ \text{or}\ b>0\\ 0,\ \text{otherwise}
\end{array}\right.
\end{align}

In this optimal control law, \eqref{fully-distri_1}-\eqref{fully-distri_2} are the expressions of the control orders for generators and LCC-HVDC systems, where $\frac{1}{\alpha_i}$ and $\frac{1}{\beta_i}$ are the optimal control coefficients, and \eqref{fully-distri_3}-\eqref{fully-distri_6} are the control dynamics. During the control process, only the communications between neighboring buses are required to exchange their information of the variables $(\theta_i,\varphi_i,\lambda_i)$, and the line variables $\gamma_{ij}^+$ and $\gamma_{ij}^-$ can be iterated at both bus $i$ and bus $j$ according to \eqref{fully-distri_5}-\eqref{fully-distri_6}, thus, control law \eqref{fully-distri} is fully-distributed.


Next, we elaborate the control design rationale, i.e., the distributed control law is designed to make the closed-loop system \eqref{system_model_dae}\eqref{fully-distri} equivalent to a partial primal-dual algorithm which solves the TS-OEFC problem \eqref{ts_oefc}. Let $\tau^G = \{ \tau_i,i\in\mathcal{N}_G\}$, $\tau^D = \{ \tau_i,i\in\mathcal{N}_D\}$, $\tau^P = \{ \tau_i,i\in\mathcal{N}_P\}$, $\varphi^G = \{ \varphi_i,i\in\mathcal{N}_G\}$, $\varphi^D = \{ \varphi_i,i\in\mathcal{N}_D\}$, $\varphi^P = \{ \varphi_i,i\in\mathcal{N}_P\}$, $\lambda^G = \{ \lambda_i,i\in\mathcal{N}_G\}$, $\lambda^D = \{ \lambda_i,i\in\mathcal{N}_D\}$, $\lambda^P = \{ \lambda_i,i\in\mathcal{N}_P\}$. First, to solve the TS-OEFC problem, we minimize $L(x,y,z)$ with respect to $\omega^G$, which yields:
\begin{align}
\label{eq_18}
0=\frac{\partial L}{\partial \omega_i} = D_i (\omega_i - \tau_i),\ i\in \mathcal{N}_G 
\end{align}

From \eqref{eq_18}, we have $\omega_i = \tau_i,i\in\mathcal{N}_G$. Substitute \eqref{eq_18} into \eqref{lag_0}, we have:
\begin{align}
\label{lag_1}
&L_1(x_1,y,z) = L(x,y,z) \big| _{\omega^G = \tau^G}
\end{align}
where $x_1 = (p^G,p^D,P,\varphi)$ denotes the primal variables, and $y$ and $z$ denote the dual variables. Then, we apply a partial primal-dual distributed algorithm \cite{feijer2010stability,wang2018distributed} to solve the TS-OEFC problem, where $L_1(x_1,y,z)$ is maximized with respect to $\tau^D$ and $\tau^P$, and thus $\tau^D$ and $\tau^P$ have no algorithm dynamic. The partial primal-dual algorithm takes the form: 
\begin{subequations}
	\label{eq_20}
	\begin{align}
	&\dot{p}_i^G=-K_i^G \frac{\partial L_1}{\partial p_i^G} = -K_i^G \left( \alpha_i(p_i^G-P_i^G)+\tau_i+\lambda_i \right) \label{eq_20_1}
	\\&\dot{p}_i^D=-K_i^D \frac{\partial L_1}{\partial p_i^D} = -K_i^D \left( \beta_i(p_i^D-P_i^D)+\tau_i+\lambda_i \right) \label{eq_20_2}
	\\&\dot{P}_{ij} = -K_{ij}^P \frac{\partial L_1}{\partial P_{ij}} = K_{ij}^P(\tau_i - \tau_j) \label{eq_20_3}
	\\&\dot{\varphi}_i = -K_{i}^{\varphi} \frac{\partial L_1}{\partial \varphi_i} = K_i^{\varphi} \bigg( \sum_{j\in \mathcal{N}_i} B_{ij}(\lambda_i - \lambda_j) \nonumber
	\\&\ \ \ \ \ \ \ \ \ \ \ \ \ \ \ \ \ \ \ \ \ \ - \sum_{e\in \mathcal{E}} C_{i,e} B_e (\gamma_e^+ - \gamma_e^-) \bigg) \label{eq_20_4}
	\\&\dot{\tau_i} = K_i^{\tau}\frac{\partial L_1}{\partial \tau_i} = K_i^{\tau} \bigg( P_i^{in} + p_i^G - D_i \tau_i - \sum_{e\in \mathcal{E}} C_{i,e} P_e \bigg),\nonumber
	\\&\ \ \ \ \ \ \ \ \ \ \ \ \ \ \ \ \ \ \ \ i\in\mathcal{N}_G \label{eq_20_5}
	\\&0=\frac{\partial L_1}{\partial \tau_i} = P_i^{in} + p_i^D - \sum_{e\in \mathcal{E}} C_{i,e} P_e,\ i\in\mathcal{N}_D \label{eq_20_6}
	\\&0=\frac{\partial L_1}{\partial \tau_i} = P_i^{in}- \sum_{e\in \mathcal{E}} C_{i,e} P_e,\ i\in\mathcal{N}_P \label{eq_20_7}
	\\&\dot{\lambda}_i = K_i^{\lambda} \frac{\partial L_1}{\partial \lambda_i} = K_i^{\lambda} \bigg( P_i^{in} + d_i p_i^G + e_i p_i^D \nonumber
	\\&\ \ \ \ \ \ \ \ \ \ \ \ \ \ \ \ \ \ \ \ - \sum_{j\in \mathcal{N}_i} B_{ij}(\varphi_i - \varphi_j)\bigg) \label{eq_20_8}
	\\&\dot{\gamma}_{ij}^+ = K_{ij}^{\gamma^+} \bigg[ \frac{\partial L_1}{\partial \gamma_{ij}^+} \bigg]_{\gamma_{ij}^+}^+ = K_{ij}^{\gamma^+} \left[B_{ij} (\varphi_i - \varphi_j)-\overline{P}_{ij} \right]_{\gamma_{ij}^+}^+  \label{eq_20_9}
	\\&\dot{\gamma}_{ij}^- = K_{ij}^{\gamma^-} \bigg[ \frac{\partial L_1}{\partial \gamma_{ij}^-} \bigg]_{\gamma_{ij}^-}^+ = K_{ij}^{\gamma^-} \left[\underline{P}_{ij} - B_{ij} (\varphi_i - \varphi_j) \right]_{\gamma_{ij}^-}^+ \label{eq_20_10}
	\end{align}
\end{subequations}
where $K_i^G$, $K_i^D$, $K_{ij}^P$ and $K_i^{\tau}$ are the stepsizes, $e_i = 1$ for $i\in \mathcal{N}_D$ while $e_i = 0$ for $i\in \mathcal{N}_G \cup \mathcal{N}_P$. By identifying $\tau_i$ with $\omega_i$, and selecting the stepsizes $K_i^G = \frac{1}{\alpha_i T_i^G}$, $K_i^D = \frac{1}{\beta_i T_i^D}$, $K_{ij}^P = B_{ij}$, $K_i^{\tau} = \frac{1}{M_i}$, \eqref{eq_20_1}-\eqref{eq_20_3} and \eqref{eq_20_5}-\eqref{eq_20_7} are identical to the system state model \eqref{system_model_dae}. \eqref{fully-distri_1}-\eqref{fully-distri_2} can be obtained by comparing \eqref{eq_20_1}-\eqref{eq_20_2} and \eqref{system_model_dae_5}-\eqref{system_model_dae_6}. Moreover, when an emergency fault occurs in the MIDC system, to avoid redundant measurements of the changes of the power injections $P_i^{in}$ in \eqref{eq_20_8}, we substitute \eqref{system_model_dae_2}-\eqref{system_model_dae_4} and \eqref{p_e_ij} into \eqref{eq_20_8}, which yields \eqref{fully-distri_3}. Therefore, \eqref{eq_20_4} and \eqref{eq_20_8}-\eqref{eq_20_10} are identical to the designed control dynamics \eqref{fully-distri}. 

Based on the above analysis, the closed-loop system \eqref{system_model_dae}\eqref{fully-distri} is equivalent to the partial primal-dual algorithm \eqref{eq_20}, which means that in case of emergency faults, this closed-loop system will reach the optimum of the TS-OEFC problem in a distributed algorithm approach. In Section IV.A, we rigorously prove the optimality of the closed-loop equilibrium. Further, according to Theorem 1, the designed optimal fully-distributed control law can guarantee the transient stability constraints and restore the system frequency.   

\subsection{Semi-Distributed Control Law}

In this subsection, we expound how to transform the fully-distributed control law \eqref{fully-distri} into a semi-distributed version. Consistent with the design rationale of the fully-distributed control, we expect that the closed-loop system with the semi-distributed control is equivalent to a new partial primal-dual algorithm solving the TS-OEFC problem. Different from \eqref{eq_20}, in the new partial primal-dual algorithm for semi-distributed design, we minimize $L_1(x_1,y,z)$ with respect to $\varphi_G$ and $\varphi_P$, and maximize $L_1(x_1,y,z)$ with respect to $\lambda_G$ and $\lambda_P$; thus, for $i\in\mathcal{N}_G \cup\mathcal{N}_P$, we have:
\begin{subequations}
	\begin{align}
	&0 = \frac{\partial L_1}{\partial \varphi_i} = -\sum_{j\in \mathcal{N}_i} B_{ij}(\lambda_i - \lambda_j) + \sum_{e\in \mathcal{E}} C_{i,e} B_e (\gamma_e^+ - \gamma_e^-) \label{eq_22_1}
	\\&0 = \frac{\partial L_1}{\partial \lambda_i} =  P_i^{in} + d_i p_i^G - \sum_{j\in \mathcal{N}_i} B_{ij}(\varphi_i - \varphi_j) \label{eq_22_2}
	\end{align}
\end{subequations}
The other parts of the algorithm for semi-distributed design are the same as \eqref{eq_20}. Then, substitute \eqref{system_model_dae_2} and \eqref{system_model_dae_4} into \eqref{eq_22_2}, we have:
\begin{align}
\label{eq_23}
&d_i (M_i \dot{\omega_i}+D_i \omega_i) + \sum_{j\in \mathcal{N}_i} B_{ij}(\theta_i - \theta_j)\nonumber
\\&-\sum_{j\in \mathcal{N}_i} B_{ij}(\varphi_i - \varphi_j) = 0,\ i\in\mathcal{N}_G \cup\mathcal{N}_P
\end{align}

Based on the above new algorithm, considering the control center of the AC main system, the optimal semi-distributed control law is presented as follows: 1) The LCC-HVDC connected buses in $\mathcal{N}_D$ conduct the same distributed control law as \eqref{fully-distri_2}-\eqref{fully-distri_4}, where the required neighboring information is from the control center. 2) The control center of the AC main system gathers the information of buses in $\mathcal{N}_G \cup \mathcal{N}_P$, receives the neighboring information from the LCC-HVDC connected buses, solves the linear equations \eqref{linear_equ} in a centralized way, and gives control orders to generators according to \eqref{fully-distri_1}. 3) All the LCC-HVDC connected buses and control center can iterate the related line variables $\gamma_{ij}^+$ and $\gamma_{ij}^-$ according to \eqref{fully-distri_5}-\eqref{fully-distri_6}.

The linear equations \eqref{linear_equ} for $i\in \mathcal{N}_G \cup\mathcal{N}_P$ can be derived by \eqref{eq_22_1} and \eqref{eq_23}, which are: 
\begin{subequations}
	\label{linear_equ}
	\begin{align}
	&\sum_{j\in \mathcal{N}_i} B_{ij}(\lambda_i - \lambda_j) = \sum_{e\in \mathcal{E}} C_{i,e} B_e (\gamma_e^+ - \gamma_e^-) \label{linear_equ_1}
	\\&\sum_{j\in \mathcal{N}_i} B_{ij}(\varphi_i - \varphi_j) = d_i (M_i \dot{\omega_i}+D_i \omega_i) + \sum_{j\in \mathcal{N}_i} B_{ij}(\theta_i - \theta_j) \label{linear_equ_2}
	\end{align}
\end{subequations}
\eqref{linear_equ_1} and \eqref{linear_equ_2} are two equations that can be solved independently. At the left sides of \eqref{linear_equ_1} and \eqref{linear_equ_2}, $\lambda^G$, $\lambda^P$ and $\varphi^G$, $\varphi^P$ are the unknown variables, while $\lambda^D$ and $\varphi^D$ are the transmitted variables from the LCC-HVDC connected buses. All the variables at the right sides of \eqref{linear_equ_1} and \eqref{linear_equ_2} are the measured variables or the locally iterated variables. Therefore, for each \eqref{linear_equ_1} and \eqref{linear_equ_2}, there are $(n_G+n_P)$ linear equations and $(n_G+n_P)$ unknown variables; moreover, it is easy to verify that the coefficient matrix is an $(n_G+n_P)$-order principal minor of $CBC^T$ with rank $(n_G+n_P)$. Thus, the solution of the equations \eqref{linear_equ} is unique.  

From the perspective of mathematical form, the  semi-distributed control law can be derived by letting $\dot{\lambda}_i=0,\dot{\varphi}_i=0,i\in\mathcal{N}_G \cup \mathcal{N}_P$ in the fully-distributed control law \eqref{fully-distri}. And the optimal expressions of $\lambda^G$, $\lambda^P$, $\varphi^G$ and $\varphi^P$ in the semi-distributed control law can be derived analytically by solving the linear equations \eqref{linear_equ} in a centralized way, which leads to a better transient performance compared to the fully-distributed control law. In addition, since the closed-loop system with the semi-distributed control is also equivalent to a primal-dual algorithm solving the TS-OEFC problem, the semi-distributed control law can also guarantee the transient stability constraints and restore the system frequency.

\subsection{Control Implementation} 

\begin{figure}[htb]
	\centering
	\includegraphics[width=0.47\textwidth]{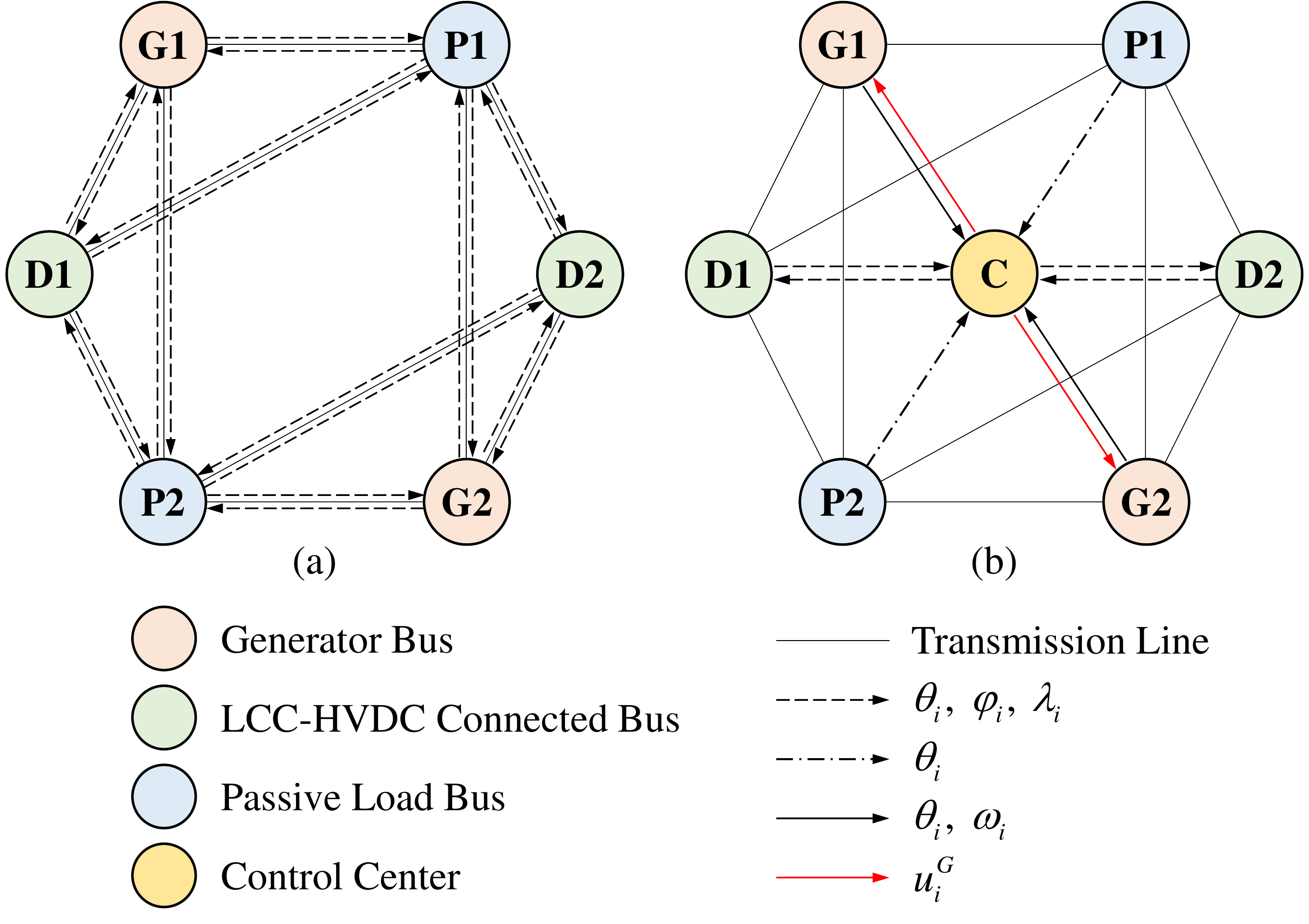}
	\caption{Communication networks of the distributed EFC strategy. (a) fully-distributed control law. (b) semi-distributed control law.}
	\label{ctrl_commu}
\end{figure}

To implement the designed distributed EFC strategy in the MIDC system, we first analyze and compare the communication networks of the fully-distributed control law and the semi-distributed control law. The communication networks for a 6-bus example system are shown in Fig. \ref{ctrl_commu}, where all types of buses, transmission lines, communication lines are displayed. According to the designed control laws in Section III.C and Section III.D, the variable information transmitted by each communication line is also shown in Fig. \ref{ctrl_commu}. The fully-distributed control law requires the communications between neighboring buses, while the semi-distributed control law requires the communications between the control center and each bus. Assuming that the communication burden of a control law can be measured by the number of the required communication lines, according to Fig. \ref{ctrl_commu}, the number of communication lines in fully-distributed scenario is $2 n_e$, while that in semi-distributed scenario is $(2n-n_P)$. Thus, for a complex power grid with numerous transmission lines, the inequality $n_e> n-\frac{1}{2}n_P$ usually holds, and the semi-distributed control law has less communication burden. 

\begin{figure}[tb]
	\centering
	\includegraphics[width=0.46\textwidth]{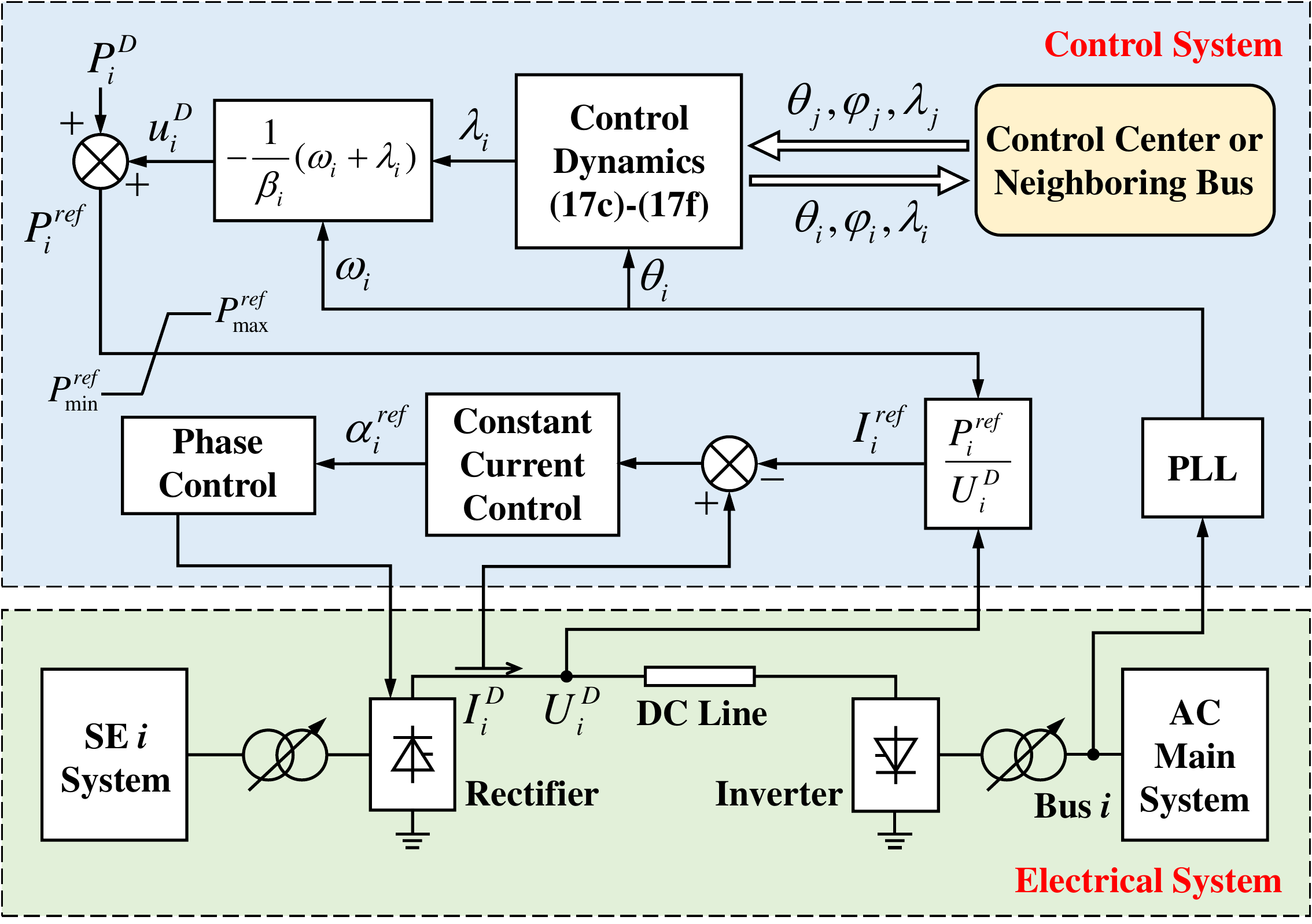}
	\caption{Control implementation in the LCC-HVDC system.}
	\label{dc_control_figure}
\end{figure}

Then, we investigate the implementation of the distributed EFC strategy in the LCC-HVDC system. Taking the SE-LCC system as an example, the control implementation is shown in Fig. \ref{dc_control_figure}. We have a similar implementation for the RE-LCC system. In Fig. \ref{dc_control_figure}, by measuring the phase angle $\theta_i$ and frequency deviation $\omega_i$ at the inverter-side AC bus $i$ (LCC-HVDC connected bus), and communicating $(\theta_i,\varphi_i,\lambda_i)$ with the control center (in semi-distributed scenario) or neighboring buses (in fully-distributed scenario), the distributed EFC strategy provides the control order $u_i^D$ for the LCC-HVDC $i$ system. Then, the DC power reference $P_i^{ref} = P_i^D+u_i^D$ is transmitted to the LCC-HVDC bipolar control station which is usually located at the rectifier side. Considering the control mode of the LCC-HVDC system, the rectifier usually adopts the constant power control or constant current control, and the DC current reference can be represented as $I_i^{ref} =P_i^{ref}/U_i^D $, where $U_i^D$ is the DC voltage \cite{arrillaga1998high,kimbark1971direct}. After the LCC-HVDC control, the signal of the phase control is output to the electrical system to regulate the power of the LCC-HVDC system.    

Moreover, we show that the designed distributed EFC can be easily integrated with the existing primary frequency control and the droop-based EFC proposed in \cite{liu2021optimal}. Let $k_i^G = \frac{1}{\alpha_i}$ and $k_i^D = \frac{1}{\beta_i}$, the first items at the right sides of \eqref{fully-distri_1} and \eqref{fully-distri_2} can be represented as $-k_i^G \omega_i$ and $- k_i^D \omega_i$, which are exactly the expressions of the droop control laws in the existing primary frequency control and the droop-based EFC. Therefore, the distributed EFC strategy contains droop-like items, and can be easily integrated with the above frequency control strategies through appropriate modifications.   

\section{Properties Analysis}

In this section, we rigorously prove the optimality of the closed-loop equilibrium and the asymptotic stability of the closed-loop system. Since the distributed EFC strategy contains two control laws and the fully-distributed control law has more complex dynamics than the semi-distributed one, we first prove the above properties of the closed-loop system with fully-distributed control law, and then we elaborate that similar results can be derived for the closed-loop system with semi-distributed control law.

\subsection{Optimality}

Since the TS-OEFC problem \eqref{ts_oefc} is strictly convex and all the constraints in \eqref{ts_oefc} are linear, the Slater's condition holds, and the KKT conditions can be utilized to characterize the primal-dual optimum. Therefore, a point $(x^*,y^*,z^*)=(p^{G*},p^{D*},\omega^{G*},P^*,\varphi^*,\tau^*,\lambda^*,\gamma^{+*},\gamma^{-*})$ is the primal-dual optimum of the TS-OEFC problem if this point satisfies the following KKT conditions: 
\begin{subequations}
	\label{kkt_con}
	\begin{align}
	&\frac{\partial L}{\partial x}\big| _{(x^*,y^*,z^*)}=0 \label{kkt_con_1}
	\\&\gamma_{ij}^{+*} (B_{ij}(\varphi_i^* - \varphi_j^*)-\overline{P}_{ij})=0,\ ij\in\mathcal{E} \label{kkt_con_2}
	\\ &\gamma_{ij}^{-*} (\underline{P}_{ij}-B_{ij}(\varphi_i^* - \varphi_j^*))=0,\ ij\in\mathcal{E} \label{kkt_con_3}
	\\ &\text{\eqref{ts_oefc_2}-\eqref{ts_oefc_9}},\ \gamma_{ij}^{+*} \ge 0,\ \gamma_{ij}^{-*} \ge 0,\ ij\in\mathcal{E} \label{kkt_con_4}
	\end{align}
\end{subequations}
where \eqref{kkt_con_1}, \eqref{kkt_con_2}-\eqref{kkt_con_3} and \eqref{kkt_con_4} represent the stationarity, complementary slackness and primal-dual feasibility conditions, respectively. Then, for the optimality of the closed-loop equilibrium with the fully-distributed control law, we have the following theorem.

\textbf{Theorem 2:} If $(p^{G*},p^{D*},\omega^{G*},P^*,\varphi^*,\lambda^*,\gamma^{+*},\gamma^{-*})$ is an equilibrium of the closed-loop system \eqref{system_model_dae}\eqref{fully-distri}, let $\tau_i = \omega_i, i\in\mathcal{N}$, then the point $(x^*,y^*,z^*)$ is the primal-dual optimum of the TS-OEFC problem.
\begin{proof}
	To prove Theorem 2, we only need to show that the KKT conditions \eqref{kkt_con} hold at the closed-loop equilibrium. The derivatives of all the state variables with respect to time equal to zero at the equilibrium. Since $\dot{p}_i^G=0$ and $\dot{p}_i^D=0$, combining \eqref{system_model_dae_5}-\eqref{system_model_dae_6} and \eqref{fully-distri_1}-\eqref{fully-distri_2}, we have:
	\begin{align}
	\label{eq_26}
	\frac{\partial L}{\partial p_i^G}| _{(x^*,y^*,z^*)}=0,\ \frac{\partial L}{\partial p_i^D}| _{(x^*,y^*,z^*)}=0
	\end{align}
	Then, due to $\dot{P}_{ij}=0$, from \eqref{system_model_dae_1_1}, we have $\omega_i^* = \omega_j^*, ij\in\mathcal{E}$; due to $\tau_i = \omega_i$, we have $\tau_i^* = \omega_i^*, i\in\mathcal{N}$. Thus:
	\begin{align}
	\label{eq_27}
	\frac{\partial L}{\partial P_{ij}}| _{(x^*,y^*,z^*)}=0,\ \frac{\partial L}{\partial \omega_i}| _{(x^*,y^*,z^*)}=0
	\end{align} 
	Since $\dot{\varphi}_i = 0$, from \eqref{fully-distri_4}, we have:
	\begin{align}
	\label{eq_28}
	\frac{\partial L}{\partial \varphi_i}| _{(x^*,y^*,z^*)}=0
	\end{align}
	Equations \eqref{eq_26}-\eqref{eq_28} show that the stationarity conditions \eqref{kkt_con_1} are satisfied at the closed-loop equilibrium. 
	
	According to the definition in \eqref{define_ba}, \eqref{fully-distri_5} and \eqref{fully-distri_6} have the same form: $\dot{a}=k [b]_a^+$, where $k$ is a positive coefficient. It is easy to verify that $a \ge 0$ is always satisfied as long as $a$ starts from a non-negative initial value. If $\dot{a}=0$, we have $[b^*]_{a^*}^+ = 0$. There are two scenarios where $[b^*]_{a^*}^+ = 0$. 1) $b^*=0$. 2) $b^* \neq 0$, $a^* \le 0$ and $b^* <0$, then due to $a^* \ge 0$, we have $a^*=0$ in this scenario. Thus, in both of the above scenarios, $b^* \le 0$ and $a^*b^*=0$ hold. Based on the above analysis, at the closed-loop equilibrium, since $\dot{\gamma}_{ij}^+ =0$ and $\dot{\gamma}_{ij}^- = 0$, the complementary slackness conditions \eqref{kkt_con_2}-\eqref{kkt_con_3} hold, the constraints \eqref{ts_oefc_8}-\eqref{ts_oefc_9} hold, and $\gamma_{ij}^{+*} \ge 0$, $\gamma_{ij}^{-*} \ge 0$. Moreover, when $\dot{\omega}_i=0$, $\dot{\lambda}_i = 0$, \eqref{system_model_dae_2}-\eqref{system_model_dae_4} and \eqref{fully-distri_3} can yield the constraints \eqref{ts_oefc_2}-\eqref{ts_oefc_7}. Therefore, the primal-dual feasibility conditions also hold at the equilibrium. In summary, the KKT conditions \eqref{kkt_con} hold at the closed-loop equilibrium, which completes this proof.  	 
\end{proof}

Theorem 2 shows the optimality of the closed-loop equilibrium with the fully-distributed control law. Since in mathematical form, the semi-distributed control law can be represented as the fully-distributed control law with $\dot{\lambda}_i=0,\dot{\varphi}_i=0,i\in\mathcal{N}_G \cup \mathcal{N}_P$, thus, for the closed-loop equilibrium with semi-distributed control law $(p^{G*},p^{D*},\omega^{G*},P^*,\varphi^{D*},\lambda^{D*},\gamma^{+*},\gamma^{-*})$, let $\tau_i = \omega_i, i\in\mathcal{N}$ and $(\lambda^G,\lambda^P,\varphi^G,\varphi^P)$ be the solution of \eqref{linear_equ}, we can also prove that $(x^*,y^*,z^*)$ is the primal-dual optimum of the TS-OEFC problem by the same idea as Theorem 2.   

\subsection{Asymptotic Stability}

In this subsection, we analyze the asymptotic stability of the closed-loop system by the Lyapunov approach. Due to the dead zone setting of the distributed EFC strategy, the post-fault initial state of the system is different from the state where the EFC strategy is enabled. We ignore the effect of the dead zone in the stability analysis, and discuss it in the case study. Moreover, due to the discontinuous dynamics introduced by the projection $[b]_a^+$ in \eqref{fully-distri_5}-\eqref{fully-distri_6}, the classic LaSalle's invariance principle is inapplicable. To solve this problem, an invariance principle for Carath\'eodory systems \cite{bacciotti2006nonpathological} is adopted in this paper. Considering the asymptotic stability of the closed-loop system with the fully-distributed control law, we have the following theorem. 

\textbf{Theorem 3:} Let $\Omega_E$ be the set of equilibriums of the closed-loop system \eqref{system_model_dae}\eqref{fully-distri}. There exists a domain $\Psi\subseteq \mathbb{R}^{2n_G} \times \mathbb{R}^{n_D} \times \mathbb{R}^{2n} \times \mathbb{R}^{3n_e}$ such that from any initial state $(p^{G0},p^{D0},\omega^{G0},P^0,\varphi^0,\lambda^0,\gamma^{+0},\gamma^{-0})\in \Psi$, the system trajectory converges to an equilibrium in $\Omega_E$.  
\begin{proof}
	Since the closed-loop system \eqref{system_model_dae}\eqref{fully-distri} is equivalent to the partial primal-dual algorithm \eqref{eq_20}, to prove Theorem 3, we only need to prove the asymptotic stability of the algorithm dynamics \eqref{eq_20}, where $\omega_i$ is identified with $\tau_i$ for $i\in\mathcal{N}$. Let $x_1=(p^G,p^D,P,\varphi)$, $y_1=(\tau^G,\lambda)$ and $z=(\gamma^+,\gamma^-)$. Denote the parameter matrices:
	\begin{align}
	&K^{x_1} = diag((T_i^G)_{i\in\mathcal{N}_G},(T_i^D)_{i\in\mathcal{N}_D},(B_{ij})_{ij\in\mathcal{E}},(\frac{1}{K_i^{\varphi}})_{i\in\mathcal{N}})\nonumber
	\\ &K^{y_1} = diag((M_i)_{i\in\mathcal{N_G}},(\frac{1}{K_i^{\lambda}})_{i\in\mathcal{N}})\nonumber
	\\&K^z = diag((\frac{1}{K_{ij}^{\gamma^+}})_{ij\in\mathcal{E}},(\frac{1}{K_{ij}^{\gamma^-}})_{ij\in\mathcal{E}})
	\end{align}
	We consider the following Lyapunov function candidate:
	\begin{align}
	\label{eq_30}
	&V(x_1,y_1,z) = \frac{1}{2} (x_1-x_1^*)^T K^{x_1} (x_1-x_1^*)\\
	&+\frac{1}{2} (y_1-y_1^*)^T K^{y_1} (y_1-y_1^*)+\frac{1}{2} (z-z^*)^T K^{z} (z-z^*)\nonumber 
	\end{align}
	where $(x_1^*,y_1^*,z^*)$ is an equilibrium of the closed-loop system, and $V$ is positive definite. 
	
	Taking the derivative of $V$ with respect to time, and combining \eqref{eq_20}, we have:
	\begin{align}
	\label{eq_31}
	\dot{V} = &-(x_1 - x_1^*)^T \frac{\partial L_1(x_1,y,z)}{\partial x_1} + (y_1 - y_1^*)^T \frac{\partial L_1(x_1,y,z)}{\partial y_1} \nonumber
	\\&+ (z - z^*)^T \bigg[\frac{\partial L_1(x_1,y,z)}{\partial z}\bigg]^+_z
	\end{align}
	For the projection $[b]_a^+$, if $a^* \ge 0$, we have: 
	\begin{align}
	(a-a^*)[b]_a^+ \le (a-a^*)b
	\label{eq_32}
	\end{align}
	which is proven as follows. According to \eqref{define_ba}, if $[b]_a^+ = b$, we have $(a-a^*)[b]_a^+ = (a-a^*)b$; if $[b]_a^+ = 0$, which means $a\le 0$ and $b\le 0$, since $a^* \ge 0$, we have $(a-a^*)b \le 0=(a-a^*)[b]_a^+$. Therefore, \eqref{eq_32} holds. As mentioned before, $\gamma_{ij}^{+*} \ge 0$, $\gamma_{ij}^{-*} \ge 0$, thus relation \eqref{eq_32} can be applied to the third item of \eqref{eq_31}. Moreover, according to \eqref{eq_20_6}-\eqref{eq_20_7}, $(\partial L_1 /\partial \tau^D) = 0$, $(\partial L_1 /\partial \tau^P) = 0$, thus, we have:
	\begin{align}
	&(y-y^*)^T \frac{\partial L_1}{\partial y} = (y_1 - y_1^*)^T \frac{\partial L_1}{\partial y_1} \label{eq_33}
	\end{align}
	According to \eqref{eq_32} and \eqref{eq_33}, letting $w=(y,z)$, we have:
	\begin{align}
	\dot{V} \le &(x_1^* - x_1)^T \frac{\partial L_1(x_1,y,z)}{\partial x_1} + (y - y^*)^T \frac{\partial L_1(x_1,y,z)}{\partial y} \nonumber
	\\&+ (z - z^*)^T \frac{\partial L_1(x_1,y,z)}{\partial z}\nonumber
	\\=&(x_1^* - x_1)^T \frac{\partial L_1(x_1,w)}{\partial x_1} + (w - w^*)^T \frac{\partial L_1(x_1,w)}{\partial w}
	\end{align}
	From \eqref{lag_0} and \eqref{lag_1}, it is easy to verify that $L_1$ is convex with respect to $x$ and concave with respect to $w$. Thus:
	\begin{align}
	&\dot{V} \le L_1(x_1^*,w)-L_1(x_1,w)+L_1(x_1,w)-L_1(x_1,w^*)\nonumber
	\\ &= \big[L_1(x_1^*,w)-L_1(x_1^*,w^*)\big]+\big[L_1(x_1^*,w^*)-L_1(x_1,w^*)\big]\label{eq_35}
	\end{align}
	Since the equilibrium $(x_1^*,w^*)$ is a saddle point of $L_1$, these two items in \eqref{eq_35} are both non-positive. Thus, we have $\dot{V} \le 0$. Moreover, since $V(x_1,y_1,z)$ is radially unbounded, the system trajectory is bounded. It follows from the invariance principle for Carath\'eodory systems in \cite{bacciotti2006nonpathological} that the system trajectory $(x_1(t),y_1(t),z(t))$ converges to the largest invariance set $\Omega_I$ which satisfies $\dot{V}=0$ between transitions of the projection $[b]_a^+$ in \eqref{eq_20_9}-\eqref{eq_20_10}, i.e.:
	\begin{align}
	&(x_1(t),y_1(t),z(t)) \rightarrow \Omega_I  \nonumber\\\Omega_I \subseteq \big\{ (x_1,y_1,&z)\big\vert \dot{V}(x_1(t),y_1(t),z(t))=0, t\notin T_p \big\}
	\end{align}  
	where $T_p = \{t_k^p,k\in \mathbb{N}\}$ is the set of projection transition time.
	
	Next, we show that any trajectory $(x_1(t),y_1(t),z(t)) \in \Omega_I$ is an equilibrium in $\Omega_E$, i.e., $\Omega_I \subseteq \Omega_E$. To have $\dot{V}=0$, both items in \eqref{eq_35} should equal to zero, i.e.:
	\begin{subequations}
	\begin{align}
	L_1(x_1,w^*)&=L_1(x_1^*,w^*) \label{eq_37_1}
	\\L_1(x_1^*,w)&=L_1(x_1^*,w^*) \label{eq_37_2}
	\end{align}
	\end{subequations}
	Differentiating \eqref{eq_37_1} with respect to time gives:
	\begin{align}
	\big(\frac{\partial L_1(x_1,w^*)}{\partial x_1}\big)^T\dot{x}_1 =0 = -\dot{x}_1^T K^{x_1} \dot{x}_1
	\end{align}
	Thus $\dot{x}_1=0$. Similarly, differentiating \eqref{eq_37_2} with respect to time gives:
	\begin{align}
	\label{eq_39}
	&\big(\frac{\partial L_1(x_1^*,w)}{\partial y_1}\big)^T\dot{y}_1 + \big(\frac{\partial L_1(x_1^*,w)}{\partial z}\big)^T\dot{z}=0
	\\ &=\dot{y}_1^T K^{y_1} \dot{y}_1+ \big(\frac{\partial L_1(x_1^*,w)}{\partial z}\big)^T (K^z)^{-1} \big[\frac{\partial L_1(x_1^*,w)}{\partial z}\big]_z^+ \nonumber
	\end{align}
	Since both items in \eqref{eq_39} are non-negative, we have $\dot{y}_1=0$. For the second item in \eqref{eq_39}, if $[\partial L_1/\partial z]^+_z=\partial L_1/\partial z$, according to \eqref{eq_39}, we have $\partial L_1/\partial z=0$, thus $\dot{z}=0$; if $[\partial L_1/\partial z]^+_z=0$, according to \eqref{eq_20_9}-\eqref{eq_20_10}, $\dot{z}=0$ also holds. Therefore, any trajectory in $\Omega_I$ is an equilibrium. 
	
	Finally, we elaborate that the system trajectory will always converge to a single equilibrium in $\Omega_E$. Based on the aforementioned analysis, the trajectory $(x_1(t),y_1(t),z(t)) \rightarrow \Omega_I \subseteq \Omega_E$ and the trajectory is bounded, hence there exists an infinite sequence of time instants $\{ t_k\}$ such that $(x_1(t_k),y_1(t_k),z(t_k)) \rightarrow (\hat{x}_1^*,\hat{y}_1^*,\hat{z}^*) \in \Omega_E$. Then, we define $V$ with the specific equilibrium $(\hat{x}_1^*,\hat{y}_1^*,\hat{z}^*)$ according to \eqref{eq_30}. Since $V$ is non-increasing with respect to time and $V$ is quadratic, we have $V(x_1,y_1,z) \rightarrow V(\hat{x}_1^*,\hat{y}_1^*,\hat{z}^*)=0$. Due to the continuity of $V$, we have $(x_1(t),y_1(t),z(t)) \rightarrow (\hat{x}_1^*,\hat{y}_1^*,\hat{z}^*)$. Therefore, the system trajectory converges to a single equilibrium in $\Omega_E$, which completes the proof.
\end{proof}

Theorem 3 guarantees the asymptotic stability of the closed-loop system \eqref{system_model_dae}\eqref{fully-distri}. Considering the closed-loop system with the semi-distributed control law, let $x_2 = (p^G,p^D,P,\phi^D)$, $y_2 = (\tau^G,\lambda^D)$, and then a quadratic Lyapunov function candidate $V_2(x_2,y_2,z)$ similar to \eqref{eq_30} can be defined and utilized to prove the asymptotic stability of the closed-loop system with the semi-distributed control law, which follows the same idea as Theorem 3.

\section{Case Study}

\subsection{System Description}

\begin{figure}[htb]
	\centering
	\includegraphics[width=0.49\textwidth]{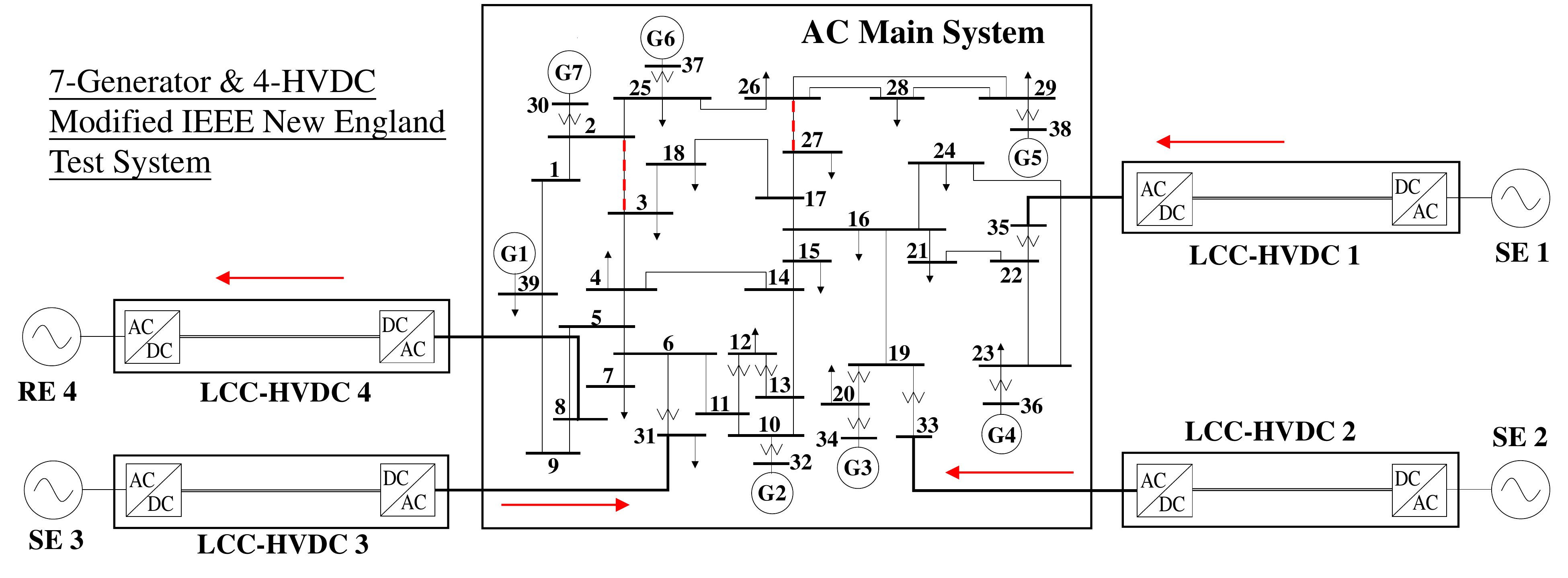}
	\caption{Topology of the MIDC test system.}
	\label{case_topo}
\end{figure}

The MIDC test system is a modified IEEE New England system combining the CIGRE LCC-HVDC system \cite{faruque2005detailed}, of which the topology is shown in Fig. \ref{case_topo}. In this test system, the AC main system contains seven generators and is connected with four LCC-HVDC systems, and LCC 1-3 are SE-LCC systems while LCC 4 is an RE-LCC system. We build the full electromagnetic transient (EMT) model of the test system on the CloudPSS platform \cite{liu2018modeling,song2020cloudpss}. In the CIGRE LCC-HVDC system, the rectifier adopts constant power control and the inverter adopts the constant extinction angle control. The adjacent AC systems adopt the equivalent centre-of-inertia (COI) model.    

To verify the effectiveness of the proposed distributed EFC strategy, the optimal fully-distributed control law and the optimal semi-distributed control law are respectively implemented in the MIDC test system. We set the active power base-value of the test system as 100 MW, and set the cost coefficients of the generators as $\alpha_1 = \alpha_5 = \alpha_6 = 0.16$ p.u. and $\alpha_2 = \alpha_3 = \alpha_4 = \alpha_7 = 0.2$ p.u.. According to \eqref{fully-distri_1}, the optimal coefficient for generator $i$ can be represented as $\frac{1}{\alpha_i}$. Set the power upper bound of LCC 1-3 as 750 MW and the power lower bound of LCC 4 as 400 MW, and the related parameters of LCC-HVDC systems are shown in Table I. Since the effective cost coefficient of LCC-HVDC $i$ is $\beta_i = \frac{2 \tilde{\beta}_i}{(Z_i^D)^2}$, according to \eqref{fully-distri_2}, the optimal coefficient for LCC-HVDC $i$ is $\frac{1}{\beta_i}$, which is also shown in Table I. We set the other control parameters as $K_i^{\lambda}=0.01$, $K_i^{\varphi}=0.01$ and $K_{ij}^{\gamma^+}=K_{ij}^{\gamma^-}=0.005$. Moreover, we assume that transmission line 2-3 and line 26-27 (red dash lines in Fig. \ref{case_topo}) have transient stability constraints, with $\underline{P}_{2-3}=\underline{P}_{26-27}=180$ MW and $\overline{P}_{2-3}=\overline{P}_{26-27}=250$ MW. 

Considering the emergency fault setting, we set a $-500$ MW power imbalance to Bus 18 at the time of 20 s, and the frequency drop of the AC main system in the following simulation results indicates this fault is an emergency. According to the above settings, the simulation results are shown as follows.     

\begin{table}[htbp] 
	\scriptsize
	\centering
	\vspace{-0.1cm}  %
	\setlength{\abovecaptionskip}{0.cm}
	\setlength{\belowcaptionskip}{-0.cm}
	\caption{Parameters of LCC-HVDC Systems}
	\label{tab1} 
	\begin{tabular}{cccccc} 
		\toprule 
		No. & $P_i^D$ & $Z_i^D$ & $\tilde{\beta}_i$ & $1/\beta_i$\\ 
		\midrule 
		LCC1 & 645 MW & 105 MW & 0.04 p.u. & 13.78 p.u.\\
		LCC2 & 630 MW & 120 MW & 0.04 p.u. & 18.00 p.u.\\
		LCC3 & 660 MW & 90 MW & 0.04 p.u. & 10.13 p.u.\\
		LCC4 & 500 MW & 100 MW & 0.04 p.u. & 12.50 p.u.\\  
		\bottomrule 
	\end{tabular} 
\end{table}

\subsection{Effectiveness of Distributed EFC}

\begin{figure}[tb]
	\centering
	\includegraphics[width=0.45\textwidth]{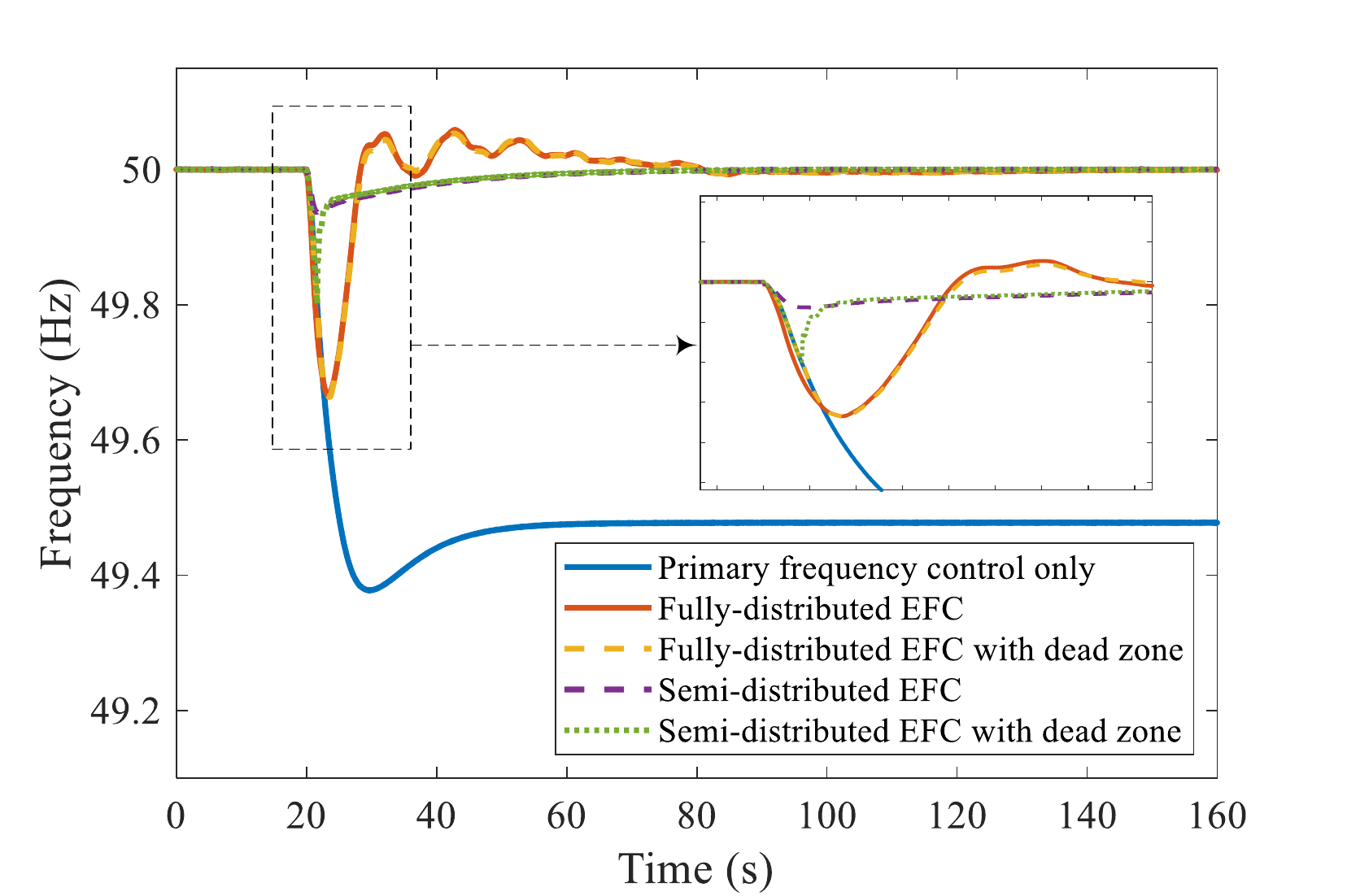}
	\caption{Frequencies of the AC main system.}
	\label{case1}
\end{figure}

\begin{figure}[tb]
	\centering
	\includegraphics[width=0.49\textwidth]{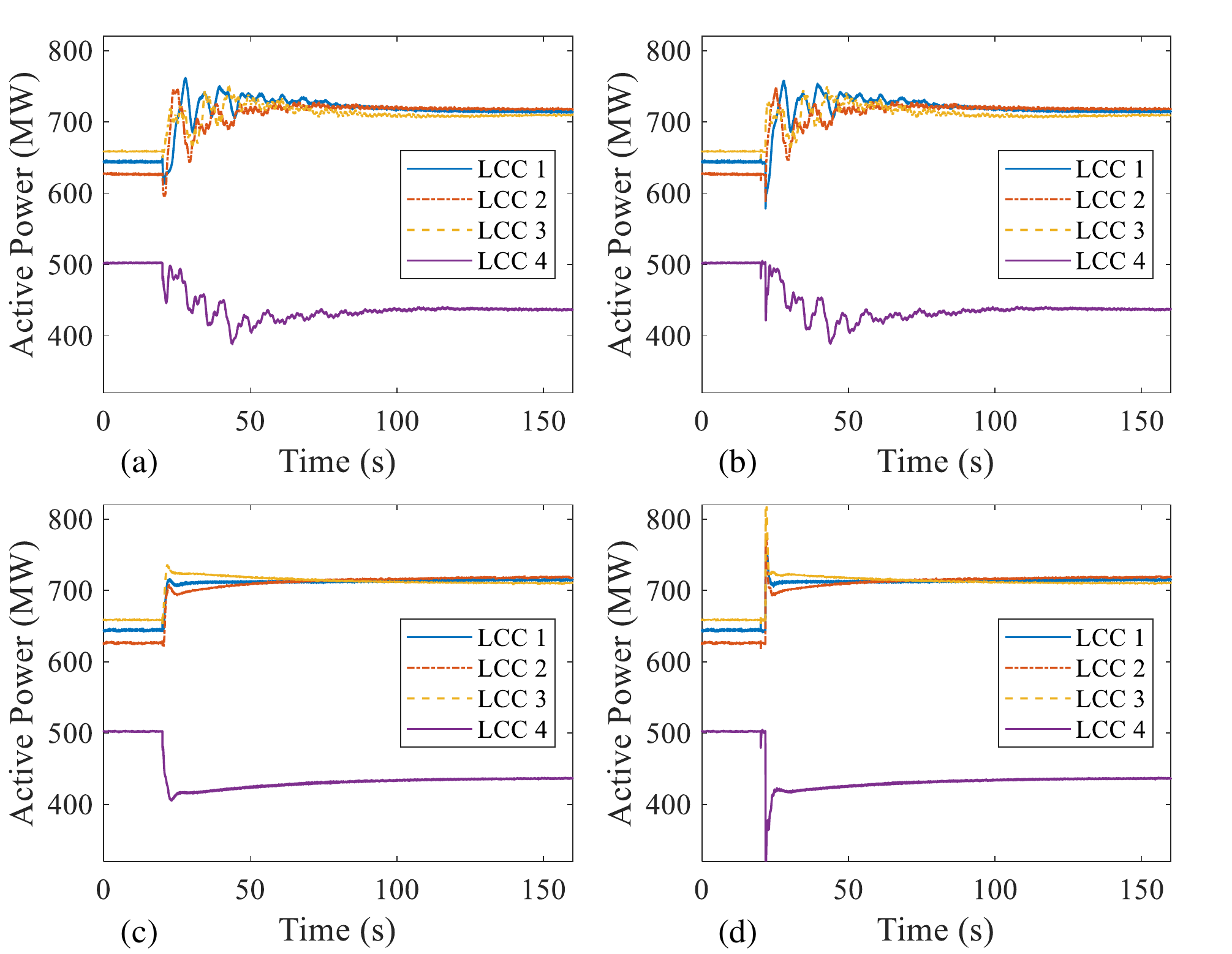}
	\caption{Active power of the LCC-HVDC systems. (a) fully-distributed EFC. (b) fully-distributed EFC with dead zone. (c) semi-distributed EFC. (d) semi-distributed EFC with dead zone.}
	\label{case2}
\end{figure}

We simulate the following five scenarios to illustrate the effectiveness of the distributed EFC: (i) The distributed EFC is not implemented, and only the generator-based primary frequency control works. (ii) The system is implemented with the fully-distributed EFC. (iii) On the basis of (ii), the EFC has dead zone. (iv) The system is implemented with the semi-distributed EFC. (v) On the basis of (iv), the EFC has dead zone. All the EFC strategies in scenario (ii)-(v) are set with the optimal coefficients. And since the grid code DL/T 428-2010 \cite{GBload} specifies that the threshold value of load shedding is supposed not to be higher than 49.5 Hz, we set the limitation value of the dead zone in scenarios (iii) and (v) to 49.8 Hz. The frequencies of the AC main system in the above five scenarios are shown in Fig. \ref{case1}, and the active power of the LCC-HVDC systems in scenario (ii)-(v) is shown in Fig. \ref{case2}. 

As shown in Fig. \ref{case1}, in scenario (i), the frequency of the AC main system drops to approximately 49.45 Hz, which has already led to severe frequency problem in engineering practice. Hence, only the generator-based primary frequency control might not ensure the frequency stability of the MIDC system in emergencies. In scenario (ii)-(v), with the proposed distributed EFC strategy, the system frequency is restored to the nominal value 50 Hz at approximately 90 s. Comparing the frequencies in scenario (ii) and (iv), we can see that with the fully-distributed EFC, the system frequency has a lower nadir and has a overshoot, while the frequency with the semi-distributed EFC has no overshoot. Thus, compared to the fully-distributed EFC, the semi-distributed EFC has a better transient control performance. From scenario (ii)-(iii) and (iv)-(v), we can find that the dead zone has no effect on the steady-state frequency and has a tiny effect on the transient process of the system frequency. As shown in Fig. \ref{case2}, when the emergency fault occurs, with the designed EFC strategy, the active power of the LCC-HVDC systems are rapidly regulated to ensure the system frequency stability. From Fig. \ref{case2}(a) and Fig. \ref{case2}(c), the active power with the fully-distributed EFC converges to the steady-state value after a relatively long transient process, while the active power with the semi-distributed EFC converges rapidly and smoothly, which also shows that the semi-distributed EFC has a better transient performance. By comparing Fig. \ref{case2}(a)(c) and Fig. \ref{case2}(b)(d), we can also find that the dead zone has little effect on the control performance. The above analysis illustrates the effectiveness of the distributed EFC strategy.

\begin{figure}[htb]
	\centering
	\includegraphics[width=0.49\textwidth]{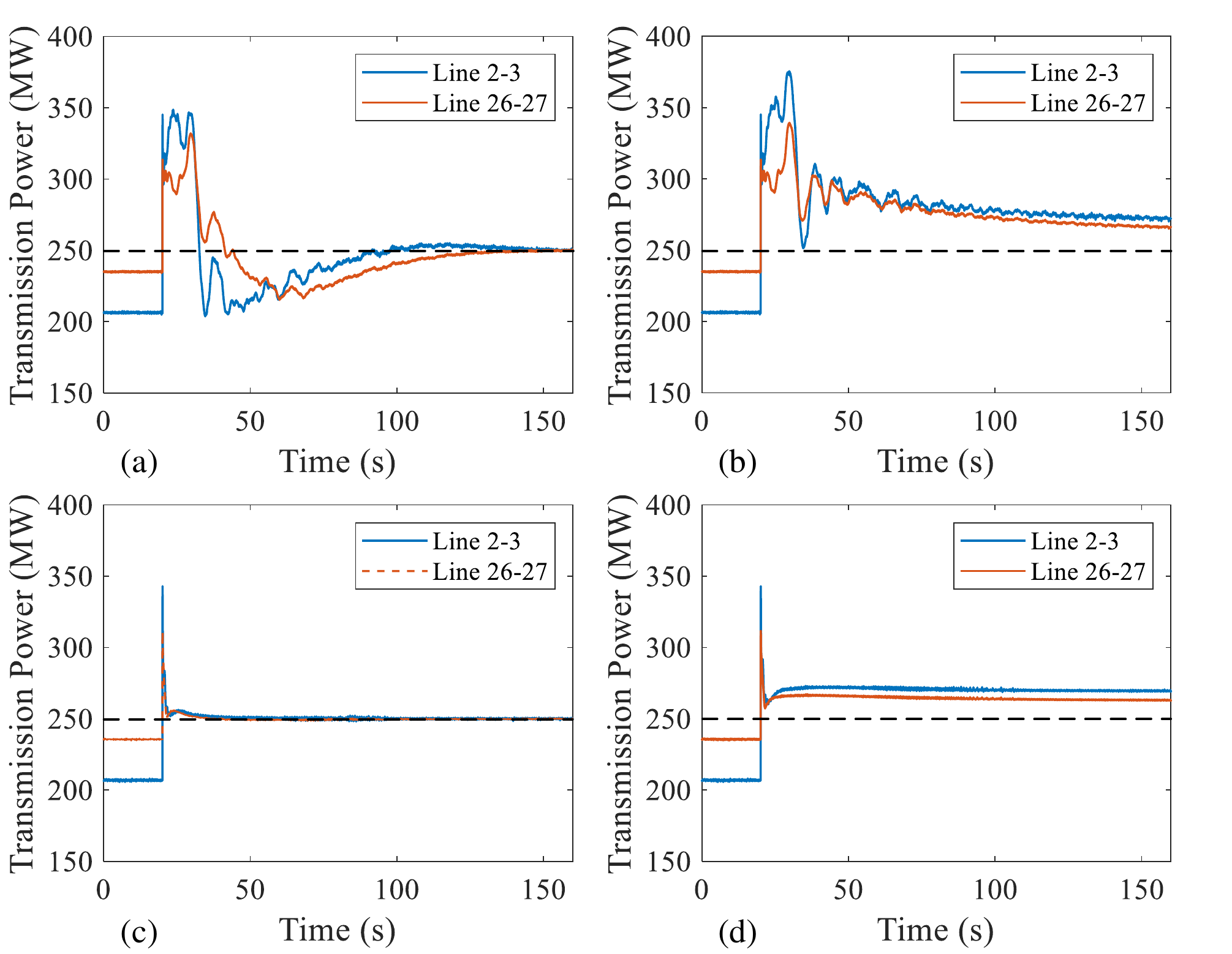}
	\caption{Transmission power of lines. (a), (b) fully-distributed EFC with / without transient stability constraints. (c), (d) semi-distributed EFC with / without transient stability constraints.}
	\label{case3}
\end{figure}

In addition, to verify that the distributed EFC guarantees the transient stability constraints of the transmission lines, the other two groups of simulations are carried out for contrast, where the transient stability constraints of line 2-3 and line 26-27 are not considered in the EFC. Fig. \ref{case3} shows the simulation results. In Fig. \ref{case3}(b)(d), the steady-state transmission power of lines exceeds the transient stability limit $\overline{P}_{2-3}=\overline{P}_{26-27}=250$ MW (black dash lines in Fig. \ref{case3}), which might lead to transient instability problems in engineering practice. And in Fig. \ref{case3}(a)(c), when the transient stability constraints are considered, the steady-state transmission power does not exceed the limit. Therefore, the simulation results in Fig. \ref{case3} shows that both the designed fully-distributed and semi-distributed EFC strategies can guarantee the transient stability constraints effectively.

\subsection{Optimality Verification}

\begin{figure}[htb]
	\centering
	\includegraphics[width=0.49\textwidth]{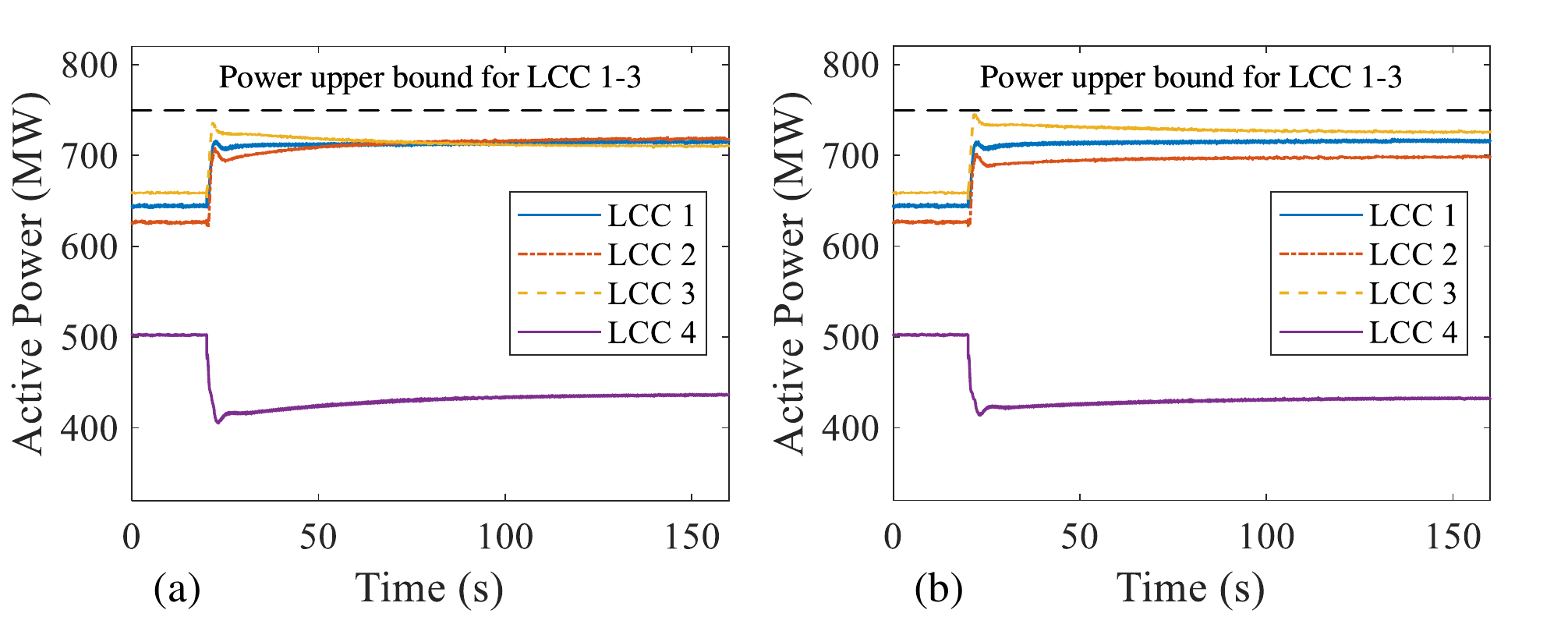}
	\caption{Active power of the LCC-HVDC systems with the semi-distributed EFC. (a) optimal coefficients. (b) average coefficients.}
	\label{case4}
\end{figure}

In Section III.B, the selected optimal control objective is to appropriately allocate the power imbalance according to the power regulation margin. In this subsection, we verify the optimality of the distributed EFC strategy. The optimal coefficients of the EFC are given in Section V.A. And for contrast, we define the average value of the optimal coefficients as the average coefficient. Thus, the average coefficients of the generators are all 5.54 p.u., and the average coefficients of the LCC-HVDC systems are all 13.60 p.u.. Two groups of simulations are carried out, which adopt the semi-distributed EFC (for example) with the optimal coefficients and the average coefficients respectively. The results are shown in Fig. \ref{case4}.

Since LCC 1-3 have the same power upper bound (black dash lines in Fig. \ref{case4}), the power regulation margins satisfy $Z_2^D > Z_1^D > Z_3^D$. In Fig. \ref{case4}(a), with the optimal coefficients, the post-fault steady-state active power of LCC 1-3 is almost same, which means that the power regulations of LCC 1-3 satisfy $u_2^D > u_1^D > u_3^D$. And in Fig. \ref{case4}(b), the power regulations of LCC 1-3 are almost equal with the average coefficients. Therefore, the optimality of the distributed EFC is verified, i.e., the LCC-HVDC system with a larger regulation margin provides more power support.   

\section{Conclusion}

In this paper, a complementary distributed EFC strategy is proposed for the MIDC system to guarantee transient stability constraints of transmission lines and restore system frequency in case of emergency faults. By selecting the semi-distributed or fully-distributed control law according to the state (normal or failure) of the control center of the AC main system, the proposed distributed EFC strategy can improve the control performance with a normal control center and enhance the control reliability in case of control center failure. Then, by formulating the optimal EFC problem with transient stability constraints (TS-OEFC), the optimal semi-distributed and fully-distributed control laws are derived by a general design rationale. Moreover, the implementation method of the distributed EFC strategy is introduced. The optimality of the closed-loop equilibrium is proven rigorously. And the asymptotic stability of the closed-loop system with the discontinuous control dynamics is proven by the Lyapunov approach and an invariance principle for Carath\'eodory systems. The simulation results of the MIDC test system on the CloudPSS platform show that the distributed EFC strategy can effectively guarantee transient stability constraints, restore system frequency and achieve the designed optimal control objective.     

\bibliographystyle{IEEEtran}
\bibliography{mybib}

\end{document}